\documentclass[aps,pra,superscriptaddress,twocolumn,amsmath,amssymb,showpacs,floatfix]{revtex4-1}
\usepackage{times}
\usepackage[utf8x]{inputenc}
\usepackage[english]{babel}
\usepackage[T1]{fontenc}
\usepackage{lmodern,braket}
\usepackage{bbold}
\usepackage[pdftex]{graphicx}
\usepackage{epstopdf}
\usepackage{bm,bbm}
\usepackage{amsmath}
\usepackage{color}
\usepackage{tikz-cd} 
\usepackage{tikz}
\usetikzlibrary{positioning}
\usepackage{adjustbox}
\usepackage{enumerate}
\usepackage[colorlinks=true,citecolor=blue,linkcolor=blue,urlcolor=blue]{hyperref}

\DeclareMathOperator{\Tr}{Tr}
\newcommand{\hs}{\mathcal{L}_2(\mathcal{H}_S)}
\newcommand{\he}{\mathcal{L}_2(\mathcal{H}_E)}
\newcommand{\hse}{\mathcal{L}_2(\mathcal{H}_S\otimes \mathcal{H}_E)}
\begin{document}
\title{Adapted projection operator technique for the treatment of initial correlations}
\author{Andrea Trevisan}
\affiliation{Dipartimento di Fisica ``Aldo Pontremoli'', Universit{\`a} degli Studi di Milano, via Celoria 16, 20133 Milan, Italy}

\author{Andrea Smirne}
\email{andrea.smirne@unimi.it}
\affiliation{Dipartimento di Fisica ``Aldo Pontremoli'', Universit{\`a} degli Studi di Milano, via Celoria 16, 20133 Milan, Italy}
\affiliation{Istituto Nazionale di Fisica Nucleare, Sezione di Milano, via Celoria 16, 20133 Milan, Italy}

\author{Nina Megier}
\affiliation{Dipartimento di Fisica ``Aldo Pontremoli'', Universit{\`a} degli Studi di Milano, via Celoria 16, 20133 Milan, Italy}
\affiliation{Istituto Nazionale di Fisica Nucleare, Sezione di Milano, via Celoria 16, 20133 Milan, Italy}

\author{Bassano Vacchini}
\email{bassano.vacchini@mi.infn.it}
\affiliation{Dipartimento di Fisica ``Aldo Pontremoli'', Universit{\`a} degli Studi di Milano, via Celoria 16, 20133 Milan, Italy}
\affiliation{Istituto Nazionale di Fisica Nucleare, Sezione di Milano, via Celoria 16, 20133 Milan, Italy}

\begin{abstract}
The standard theoretical descriptions of the dynamics of open quantum systems rely on the assumption that the correlations 
with the environment can be neglected at some reference (initial) time. 
While being reasonable in specific instances, such as when the coupling between the system and the environment is weak or
when the interaction starts at a distinguished time, the use of initially uncorrelated states is questionable
if one wants to deal with general models, taking into account the mutual influence 
that the open-system and environmental evolutions perform on each other.
Here, we introduce a perturbative method
that can be applied to any microscopic modeling of the system-environment interaction, including
fully general initial correlations. Extending the standard technique based on projection operators that single out the relevant
part of the global dynamics, we define a family of projections adapted to a convenient decomposition
of the initial state, which involves a convex mixture of product operators with proper environmental states. 
This leads us to characterize the open-system dynamics via an uncoupled system of differential equations, which are homogeneous
and whose number is limited by the dimensionality of the open system, for any kind of initial correlations.
Our method is further illustrated by means of two cases study,
for which it reproduces the expected dynamical behavior in the long-time regime more consistently than the standard projection technique.
\end{abstract}

\maketitle

\section{Introduction}
\label{sec:intro}
The realistic characterization of quantum systems interacting with an environment, i.e., open quantum systems \cite{Breuer2002,Rivas2012},
plays a key role both from the conceptual and the practical point of view, whenever one aims to 
a general understanding of quantum evolutions, possibly in view of the control of quantum properties of the physical system
at hand. The complexity of the global system composed by the open system and the environment 
calls for rather drastic simplifications, to obtain a self-contained description of the relevant degrees of freedom.
The assumption that the open system and the environment are uncorrelated at the initial time is usually the very starting
point for 
a microscopic modeling of the dynamics. Besides simplifying the equations of motion, the presence of an initial global product state
guarantees that the open-system dynamics is fixed by completely positive and trace preserving (CPTP) maps defined for a generic
initial condition, in this way providing the description of the dynamics with the rich mathematical structure of CPTP maps \cite{Heinosaari2012}.

However, the choice of an initial product state
has to be put under scrutiny, like all other assumptions used to treat open quantum systems, whenever one wants to associate a given description to concrete physical systems.
While the absence of initial system-environment correlations is naturally motivated if the interaction between the open system and the environment
starts at a specific instant of time, and it can be rigorously proven to be justified in the weak-coupling regime \cite{Tasaki2007,Yuasa2007},
it is by now clear that initial correlations can have instead a significant impact in many situations, including
the interaction of a two-level system with bosonic modes \cite{Morozov2011a,Morozov2012a,Kitajima2013a,Kitajima2017a}, 
the damped harmonic oscillator \cite{Grabert1988,Davila1997,Pollak2008,Tan2011}, 
spin systems \cite{Majeed2019}, 
or even many body \cite{Myohanen2008,Chaudhry2013}
and transport-related \cite{Pomyalov2010,Velicky2010,Ma2015,Buser2017} open-systems.
In addition, the full understanding of the role of the correlations, and possibly of their quantum or classical nature, in
the evolution of open quantum systems should indeed include the analysis of those correlations that are present between
the system and the environment at the initial time, thus complementing the related studies on the correlations built up
by the dynamics
\cite{DeSantis2019,Kolodynski2020,Banacki2020,Megier2021,Smirne2021}

As a consequence, the dynamics of open quantum systems in the presence of initial correlations with the environment 
has been the object of intense study, even though a general convenient treatment of such dynamics is still missing. Mostly, the investigation has been focused on the possibility to define reduced maps at the level of the set of states
of the open system only, and, in case, to extend the CPTP property to this scenario \cite{Pechukas1994,Alicki1995,Lindblad1996,Buzek2001,Jordan2007,Modi2012,McCracken2013,Brodutch2013,Buscemi2014,Liu2014,Vacchini2016a,Shabani2016,Schmid2019,Silva2019}. 
What is more, it was shown that
specific behaviors of distinguishability quantifiers among quantum states, which can be tomographically reconstructed, can be traced back to the presence \cite{Laine2010,Smirne2010b,Smirne2011,Dajka2011,Wissmann2013,Amato2018} 
or even to the classical or quantum nature \cite{Gessner2011,Gessner2014} of initial correlations.

On the other hand, knowing that the open-system dynamics can be described via, possibly CPTP, maps
does not mean that one is actually able to evaluate the action of these maps and thus to obtain 
explicit predictions about
physical quantities of interest. 
Perturbative techniques represent a general strategy yielding an explicit characterization of the open-system dynamics
that is approximate, but that can be applied in
principle to any model and is linked directly to the microscopic features defining the system-environment interaction.
As relevant examples, let us mention the second-order expansion in the coupling constant of the propagator 
expressed in the Bargmann coherent-state basis \cite{Halimeh2017}, 
the expansion building on the system-environment correlations and leading to coupled reduced system and environmental
integro-differential equations \cite{Alipour2020},
and the perturbative method tailored to the correlations built up by the previous system-environment
interaction \cite{Trushechkin2021}.
Furthermore, a systematic perturbative approach can be obtained by means of a cumulant expansion \cite{vanKampen1974a,vanKampen1974b} 
defined via projection operators, which single out the part of the global unitary dynamics that is relevant for the evolution of the open system.
While the standard method uses projections into product states \cite{Shibata1977,Breuer2002},
correlated-projection techniques can be defined in full generality \cite{Breuer2006,Breuer2007,Mallayya2019,Huang2020,Riera2021,Donvil2021}.

In this paper, we introduce a refined version of the projection operator techniques, which combines the standard
approach based on projections into product states with a recently introduced representation of the open-system dynamics \cite{Silva2019}.
Relying on the theory of frames \cite{Ali2000,Renes2004a}, the latter is based on the decomposition of any initial global state into a convex combination of product operators, where the operators on the environment are guaranteed to be proper states, while those on the open system are not,
so that also initial entangled states can be taken into account. Defining
a family of projectors into product states -- one for each state in the decomposition -- we derive a description of the open-system
dynamics that
always consists of a family of uncoupled homogeneous differential equation, whose number
is limited by the dimensionality of the open system and not of the environment.
In addition, we also show how the mentioned representation of the initial global state can be used in the presence
of a single projection operator to get a general, more explicit form of the resulting equations of motion and connect them
with physically-relevant environmental correlation functions.
Note that we focus on the time-local version of the projection-operator techniques, leading to (system of) differential equations,
but the latter can be linked with the time-non-local version leading to integro-differential equations \cite{Reimer2019b,Nestmann2021a,Nestmann2021b}.

After deriving the explicit form of the second-order equations for a fully general microscopic model and initial system-environment state, we consider two simple paradigmatic cases study for the open-system dynamics of a qubit; namely, pure dephasing and damping by a bosonic bath. The first model 
describes a two-level system undergoing only decoherence due to the interaction with the environment and it possesses an analytic solution, which allows us to compare our general approximated expressions with the exact result, while the second, which is not exactly solvable, includes an energy exchange between the open system and the environment. 
To the best of our knowledge this is actually the first time that states with initial correlations, that is in which the two-level system is correlated directly with the bath, are considered for this model.
We also compare the predictions of our perturbative approach to those of the standard projection operators, focusing on the intermediate and long time regime, where the two descriptions can differ significantly.

The rest of the paper is organized as follows. In Sec.\ref{sec:stand-corr-proj},
we introduce the main features of the product-state projection operator method and its correlated-state generalization
that will be useful for the following. In Sec. \ref{sec:ape}, after recalling the global-state
decomposition put forward in \cite{Silva2019} and applying it to the standard projection operator techniques,
we present the main finding of the paper, that is, the systematic definition of a perturbative expansion based
on a family of product-state operators, adapted to the decomposition of the initial system-environment state.
Our results are futher discussed by means of examples in Sec.\ref{sec:examples}, while the general
conclusions and possible outlooks of our work are given in Sec.\ref{sec:conclusions-outlook}.

\section{Time-local projection-operator techniques}\label{sec:stand-corr-proj}

The main idea behind projection operator techniques applied to open-system dynamics 
is to introduce a projection 
at the level of the overall system-environment evolution, capturing
the relevant part of the global state, that is, the one needed to reconstruct the reduced state
at a generic time \cite{Breuer2002}.
In particular, this can lead both to time-local and time-non-local, i.e., integro-differential master equations,
which can be expanded perturbatively to get an explicit characterization of the reduced dynamics. 
Importantly, the error due to the truncation
of the expansion can be estimated in full generality and can be
reduced by taking into account higher orders.
On the other hand, due to the usual complexity of the perturbative expansion
\footnote{For a systematic procedure to express all the orders of the expansion in a compact
recursive way see \cite{Gasbarri2018}}, it is desirable to get
well-behaved solutions already when restricting to the lowest orders. It is then important
to compare different expansions, based on the definition of different projections or on distinct
decompositions of the initial global state $\rho_{SE}$, to evaluate which one yields 
a better description, once we fix the order of truncation. 
Here, we consider different perturbative expansions, all of them
taking into account a possibly correlated initial state $\rho_{SE}$; moreover, we restrict our analysis
to time-local, or time-convolutionless (TCL), master equations, expanded up to the second order.

Given an open system $S$, associated with the Hilbert space $\mathcal{H}_S$,
and an environment $E$, associated with $\mathcal{H}_E$, 
let us assume that their joint
dynamics at different times $t$ is fixed by a group of unitary operators 
$U(t)$ (where we set $t_0=0$ as the initial time) on the global
Hilbert space $\mathcal{H}_S \otimes \mathcal{H}_E$, i.e., we assume that
the system and the environment together form a closed system.
The open-system state $\rho_S(t)$, also called reduced state, at a generic time $t$ is
an element of the set of statistical operators $\mathcal{S}(\mathcal{H}_S)$, i.e. the
linear operators on $\mathcal{H}_S$ that are positive and with unite trace,
and it can always
be written in terms of a map from the set of statistical operators 
on the global $S-E$ degrees of freedom $\mathcal{S}\left(\mathcal{H}_{S}\otimes \mathcal{H}_{E}\right)$ 
to $\mathcal{S}(\mathcal{H}_S)$.
This map consists in the composition of the unitary evolution
and the partial trace on the environmental degrees of freedom $\mbox{Tr}_E$, according to
\begin{equation}\label{eq:rst}
\rho_S(t) = \Tr_E \left[U(t)\rho_{SE}U(t)^{\dagger}\right],
\end{equation}
and it is CPTP, while its domain involves the whole
$\mathcal{S}\left(\mathcal{H}_{S}\otimes \mathcal{H}_{E}\right)$. 
On the other hand, when we deal
with the evolution of an open quantum system, we would like to focus our description
on maps defined on $\mathcal{S}(\mathcal{H}_S)$ only. 

To achieve this, we can introduce a projection operator $\mathcal{P}$, that is a linear map such that 
$\mathcal{P}^2=\mathcal{P}$, on the set 
of bipartite Hilbert-Schmidt operators $\hse$, 
and additionally require that
$
\Tr_E \mathcal{P} = \Tr_E,
$
so that the projection is trace-preserving and in particular preserves the reduced dynamics
\begin{equation}\label{eq:rels}
\rho_S(t) = \Tr_E\left[\rho_{SE}(t)\right]=\Tr_E\left[\mathcal{P}[\rho_{SE}(t)]\right].
\end{equation}
This relation means that the reduced state $\rho_S(t)$ at a generic time
can be obtained from the evolution of the \textit{relevant part} $\mathcal{P}[\rho_{SE}(t)]$
of the global state; in fact,
projection operator techniques define general 
procedures to get closed dynamical equations for the relevant part.
The starting point is a given 
microscopic model of the open system, the environment and their interaction, as fixed by the
global Hamiltonian (which we take for simplicity time-independent)
\begin{equation}\label{eq:h}
H = H_S \otimes \mathbb{1}_E + \mathbb{1}_S \otimes H_E + g H_I,
\end{equation}
with the three terms at the right hand side representing, respectively, the free system and environment Hamiltonians, and their interaction Hamiltonian; $g$ is a dimensionless parameter quantifying the strength
of the coupling, which will be useful for the perturbative expansions.
The evolution of the global state $\rho_{SE}(t)$ is fixed by the Liouville-von Neumann equation,
which in the 
interaction picture reads 
\begin{equation}
\label{eq:VN_rho}
\frac{d}{d t}\rho_{SE}(t)=-i[g H_{I}(t),\rho_{SE}(t)] = g \mathcal{L}(t)[\rho_{SE}(t)],
\end{equation}
where we introduced the Liouville map
$\mathcal{L}(t)[\bullet]=-i[H_{I}(t),\bullet]$ and $H_{I}(t) = e^{i H_0 t} H_I e^{-i H_0 t}$
is the interaction Hamiltonian in the interaction picture (we set $\hbar =1$). 
Now, 
applying
the projection $\mathcal{P}$ 
on both sides of Eq. \eqref{eq:VN_rho}
and introducing its complementary 
$\mathcal{Q}=\mbox{Id}_{SE}-\mathcal{P}$
(using $\mbox{Id}_{SE}$ to denote the identity map on $\hse$), along 
with the propagator forward in time of the irrelevant part of the dynamics
($T_{\leftarrow}$ is the time-ordering operator)
\begin{equation}
\label{eq:tclg1}
\mathcal{G}(t,t_1)=T_{\leftarrow}\exp\left[g \int_{t_1}^t d\tau\mathcal{Q}\mathcal{L}(\tau) \right],
\end{equation}
the propagator backward in time of the global dynamics
($T_{\rightarrow}$ is the antichronological time-ordering operator)
\begin{equation}\label{eq:tclg2}
G(t,t_1)=T_\rightarrow\exp\left[-g \int_{t_1}^t d\tau\mathcal{L}(\tau)\right],
\end{equation}
and the map
\begin{equation}
\Sigma(t)=g \int_0^t dt_1 \mathcal{G}(t,t_1)\mathcal{Q}\mathcal{L}(t_1)\mathcal{P}G(t,t_1),
\end{equation}
one can derive the following equation for the relevant part of the dynamics \cite{Breuer2002}
\begin{equation}
\label{eq:rel_part_dyn_TCL}
\frac{d}{dt}\mathcal{P}[\rho_{SE}(t)]=\mathcal{K}_{TCL}(t)\mathcal{P}[\rho_{SE}(t)]+ \mathcal{I}_{TCL}(t)\mathcal{Q}[\rho_{SE}],
\end{equation}
with
the time-local generator, called TCL generator, 
\begin{align}
\mathcal{K}_{TCL}(t)&=g \mathcal{P}\mathcal{L}(t)[\mbox{Id}_{SE}-\Sigma(t)]^{-1}\mathcal{P} \label{eq:tclg}
\end{align}
and the inhomogeneity
\begin{align}
\mathcal{I}_{TCL}(t)&=g \mathcal{P}\mathcal{L}(t)[\mbox{Id}_{SE}-\Sigma(t)]^{-1}\mathcal{G}(t,0)\mathcal{Q}.
\label{eq:II}
\end{align}
This equation is well-defined for times where the operator 
$\mbox{Id}_{SE}-\Sigma(t)$ is invertible,
which is always the case for times short enough (depending on the coupling $g$)
since $\Sigma(0)=0$ \cite{Breuer2002}. 
Under this condition, Eqs.(\ref{eq:rel_part_dyn_TCL})-(\ref{eq:II})
are equivalent to the initial Liouville-von Neumann equation (\ref{eq:VN_rho}), so that
Eq.(\ref{eq:rel_part_dyn_TCL}) is as difficult to solve as the full unitary global evolution;
on the other hand, Eq.(\ref{eq:rel_part_dyn_TCL}) is the starting point
for a systematic perturbative expansion 
of the open-system dynamics. 

\subsection{Standard projection}\label{sec:stand-corr-proj-pp}
Now, different equations, as well as different perturbative expansions, are obtained from 
Eqs.(\ref{eq:rel_part_dyn_TCL})-(\ref{eq:II}) depending on the specific choice of
$\mathcal{P}$. Within the standard projection operator approach, one considers a projection given by \cite{Breuer2002}
\begin{equation}\label{eq:pprod}
	\mathcal{P}=\Tr_E[\bullet ] \otimes \overline{\rho}_E,
\end{equation} 
where $\overline{\rho}_E$ is a reference environmental state, i.e.,
the system-environment state $\rho_{SE}$ is projected by $\mathcal{P}$ into the product state $\rho_S \otimes \overline{\rho}_E$.
Such a choice is the natural one if the initial system-environment
state is a product state, with a fixed state of the environment, i.e., $\rho_{SE}=\rho_S\otimes \rho_E$,
in which case using Eq.(\ref{eq:pprod}) with $\overline{\rho}_E=\rho_E$
would indeed make the inhomogeneous term in Eq.(\ref{eq:rel_part_dyn_TCL})
equal to zero, as $\mathcal{Q}[\rho_{SE}]=0$.
More in general, Eq.(\ref{eq:pprod}) can be used also in the presence of initial correlations,
even if in this case it is a-priori not clear which choice of the reference state $\overline{\rho}_E$
can be convenient, and other projections
that reflect the initial correlations
could be actually preferred, as will be discussed in
the following. 

Assuming that the inverse of $\mbox{Id}_{SE}-\Sigma(t)$ can be expanded into the geometric
series (which is also guaranteed for times short enough, see the remark after Eq.(\ref{eq:II})), 
$[\mbox{Id}_{SE}-\Sigma(t)]^{-1} = \sum_{n=0}^{\infty}[\Sigma(t)]^n$, 
by substituting the
expression for the projection operator given by Eq.~\eqref{eq:pprod}
into Eqs.(\ref{eq:tclg}) and (\ref{eq:II}), 
we first expand 
the propagators $\mathcal{G}(t,t_1)$ and $G(t,t_1)$ with respect
to the coupling $g$, which gives a perturbative evaluation of the relevant part of the global dynamics. 
Taking then
the partial trace over the environment in Eq.(\ref{eq:rel_part_dyn_TCL}), we
obtain the perturbative expansion on the reduced dynamics, which up to second order in $g$ 
reads
\cite{Breuer2002}
\begin{align}\label{eq:stst}
\frac{d}{dt}\rho_S(t) =& \mathcal{J}_S^{(1)}(t)[\mathcal{Q}[\rho_{SE}]] + \mathcal{J}_S^{(2)}(t)[\mathcal{Q}[\rho_{SE}]] \\
&  +\mathcal{J}_S^{(1)}(t)[\rho_S(t)\otimes\overline{\rho}_E]+ \mathcal{J}_S^{(2)}(t)[\rho_S(t)\otimes\overline{\rho}_E],\notag
\end{align}
where we have defined the maps
\begin{align}
	 \mathcal{J}_S^{(1)}(t)[\bullet]=&g \Tr_E\Big[\mathcal{L}(t) [\bullet] \Big] ; \notag\\
	 \mathcal{J}_S^{(2)}(t)[\bullet]=&g^2 \int_0^t d\tau\Bigg\{ \Tr_E\Big[\mathcal{L}(t)\mathcal{L}(\tau)[\bullet]\Big] \notag\\
	 &-\Tr_E\Big[\mathcal{L}(t)\mathcal{P}\mathcal{L}(\tau)[\bullet]\Big]\Bigg\}. \label{eq:jj}
\end{align}

\subsection{Correlated-state projection}
As second choice, we consider a much wider class of projections, namely those that are in the form
\begin{equation}
\label{eq:pos_proj_form}
\mathcal{P}=\mbox{Id}_S\otimes\Lambda,
\end{equation} 
where $\Lambda: \he \to \he$ is a CP, trace-preserving and idempotent map, 
which ensure that $\mathcal{P}^2=\mathcal{P}$, as well
as the validity of Eq.(\ref{eq:rels}). For these projections there exists a representation theorem \cite{Breuer2007} stating that they
can always be written as
\begin{equation}
\label{eq:repr_theor_proj}
\mathcal{P}[\bullet]=\sum_i \Tr_E[(\mathbb{1}_S\otimes \overline{Y}_i)\bullet]\otimes \overline{X}_i,
\end{equation}
with $\{\overline{Y}_i\}$ and $\{\overline{X}_i\}$ self-adjoint environmental operators satisfying
\begin{align}
\label{eq:corr_proj_prop_1}
&\Tr_E[\overline{X}_i \overline{Y}_j]=\delta_{ij}, \notag\\
&\sum_i \Tr_E[\overline{X}_i]\overline{Y}_i=\mathbb{1}_E, \notag\\
&\sum_i \overline{Y}_i^T\otimes \overline{X}_i\geq 0.		
\end{align}	
The standard projection defined in Eq.(\ref{eq:pprod}) is a special case
of the construction above, for a single pair of environmental operators given by 
$\overline{X}=\overline{\rho}_E$ and $\overline{Y}=\mathbb{1}_E$. 
More in general, the projection in Eq.(\ref{eq:repr_theor_proj}) implies
that the relevant part of the bipartite state $\rho_{SE}(t)$ at time $t$
takes the form
\begin{equation}\label{eq:prhosecorr}
	\mathcal{P}[\rho_{SE}(t)]=\sum_i \eta_i(t)\otimes \overline{X}_i,
\end{equation}
with
\begin{equation}\label{eq:etai}
	\eta_i(t)=\Tr_E\big[(\mathbb{1}_S\otimes \overline{Y}_i)\rho_{SE}(t)\big],
\end{equation}
which then includes system-environment correlations.
Note that the proof of the representation theorem is not constructive,
so that additional insights are necessary in order to determine a
relevant choice of operators $\{\overline{Y}_i\}$ and
$\{\overline{X}_i\}$. Indeed, up to now this has been successfully considered only
for structured environments where the coupling between system and
environment was dictated by the structure of the environment \cite{Breuer2006,Riera2021}.

Replacing Eqs.(\ref{eq:prhosecorr}) and (\ref{eq:etai}) into Eq.(\ref{eq:rel_part_dyn_TCL})
and using the first identity in Eq.(\ref{eq:corr_proj_prop_1}),
it is possible to write a dynamical equation for each component $\eta_i(t)$, as 
\begin{equation}\label{eq:corr_proj_eta_dyn_TCL}
\frac{d}{dt}\eta_i(t)=\mathcal{K}_i(t)[\mathcal{P}[\rho_{SE}(t)]]+ \mathcal{I}_i(t)[\mathcal{Q}[\rho_{SE}(0)]],
\end{equation}
with 
\begin{align}
\mathcal{K}_i(t)[\bullet]&=\Tr_E \big[(\mathbb{1}_S\otimes \overline{Y}_i)\mathcal{K}_{TCL}(t)[\bullet]\big], \nonumber\\
\mathcal{I}_i(t)[\bullet]&=\Tr_E \big[(\mathbb{1}_S\otimes \overline{Y}_i)\mathcal{I}_{TCL}(t)[\bullet]\big].
\end{align}

Expanding the exact equation (\ref{eq:corr_proj_eta_dyn_TCL})
up to second order and
generalizing the definitions in Eq.(\ref{eq:jj}) as
\begin{align}
\mathcal{J}_i^{(1)}(t)[\bullet]=& g \Tr_E\Big[(\mathbb{1}_S\otimes \overline{Y}_i)\mathcal{L}(t) [\bullet] \Big] ;\nonumber\\
\mathcal{J}_i^{(2)}(t)[\bullet]=& g^2\int_0^t d\tau\Bigg\{ \Tr_E\Big[(\mathbb{1}_S\otimes \overline{Y}_i)\mathcal{L}(t)\mathcal{L}(\tau)[\bullet]\Big] \nonumber\\
&- \Tr_E\Big[(\mathbb{1}_S\otimes \overline{Y}_i)\mathcal{L}(t)\mathcal{P}\mathcal{L}(\tau)[\bullet]\Big]\Bigg\},\label{eq:jjcorr}
\end{align}
we obtain (compare with Eq.(\ref{eq:stst}))
\begin{align}
\label{eq:dyn_Delta_corr_proj}
&\frac{d}{dt}\eta_i(t)= \mathcal{J}_i^{(1)}(t)[\mathcal{Q}[\rho_{SE}]] + \mathcal{J}_i^{(2)}(t)[\mathcal{Q}[\rho_{SE}]] \\&+\sum_{j}\left(\mathcal{J}_i^{(1)}(t)\left[\eta_{j}(t)\otimes \overline{X}_{j}\right]+ 
\mathcal{J}_i^{(2)}(t)\left[\eta_{j}(t)\otimes \overline{X}_{j}\right]\right). \notag
\end{align}

In the next section, we will see how both the expressions in Eq.(\ref{eq:stst}) and in Eq.(\ref{eq:dyn_Delta_corr_proj})
can take a more explicit form by using a proper decomposition of the initial global state, from which the reduced state $\rho_S(t)$ 
can thus be obtained.

\section{Adapted perturbative expansions}\label{sec:ape}
After recalling the general formalism of projection-operator techniques, 
  we will now introduce a novel projection-operator expansion based on
  a decomposition of the initial system environment state in terms of
  positive environmental operators, rather than on a decomposition of
  the projection operator as in Eq.(\ref{eq:repr_theor_proj}). Since
  this new expansion is specifically tailored
to a representation of the initial correlated state
as a convex mixture of tensor-product operators with positive
environmental states, we will call it \emph{adapted projection
  operators} (APO) technique.
As we will show, this representation of the initial state directly
follows from the expression of the state itself, at variance with the
representation of correlated projection operators that has to be
introduced on the basis of some additional information.
Before establishing the APO technique, we will show that also the standard expansions can take advantage of such a decomposition, so as to make the comparison between the two approaches easier.

\subsection{Decomposition of bipartite states via positive environmental operators}
\label{sec:one-sided-positive}

Every bipartite statistical operator
$\rho_{SE}\in\mathcal{S}(\mathcal{H}_{S}\otimes\mathcal{H}_{E})$ can be written as \cite{Silva2019}
\begin{equation}\label{eq:opd}
\rho_{SE}= \sum_{\alpha=1}^{\mathfrak{N}} \omega_{\alpha} D_{\alpha}\otimes \rho_{\alpha},
\end{equation}
where the $\rho_{\alpha} \in \mathcal{S}(\mathcal{H}_E)$ are statistical operators on the environment and
the $\omega_\alpha > 0$ are positive numbers, while the $D_{\alpha}$ are operators 
within the set $\hs$ of Hilbert-Schmidt operators on
$\mathcal{H}_S$, i.e., the trace of the square of their absolute value is finite, 
but they are not necessarily positive.
If the $D_{\alpha}$ are also positive operators, the state $\rho_{SE}$ in Eq.(\ref{eq:opd})
is a separable state \cite{Bengtsson2006}, and if in addition the $D_{\alpha}$ or the $\rho_{\alpha}$ or both
are given by a family of orthogonal projections, $\rho_{SE}$ is a zero discord state \cite{Ollivier2001,Henderson2001,Modi2012} (according
to, respectively, the asymmetric or the symmetric definitions for bipartite states). 
Nevertheless, we stress once
more that every bipartite state, including any kind of classical or quantum correlations, 
possesses a decomposition as in Eq.(\ref{eq:opd}).
Such a decomposition can be constructed explicitly by means of
frame theory \cite{Ali2000,Renes2004a}, which also allows one to connect in full generality 
the number $\mathfrak{N}$ of terms with
the rank of $\rho_{SE}$  \cite{Smirne2021b}.
This implies that $\mathfrak{N}$ is limited by the dimensionality $d$ of the reduced system, 
being anyway bounded by $\mathfrak{N}\leq d^2$, for any dimensionality
of the environment and initial system-environment correlations.

\begin{figure*}
 \includegraphics[width=1.9\columnwidth]{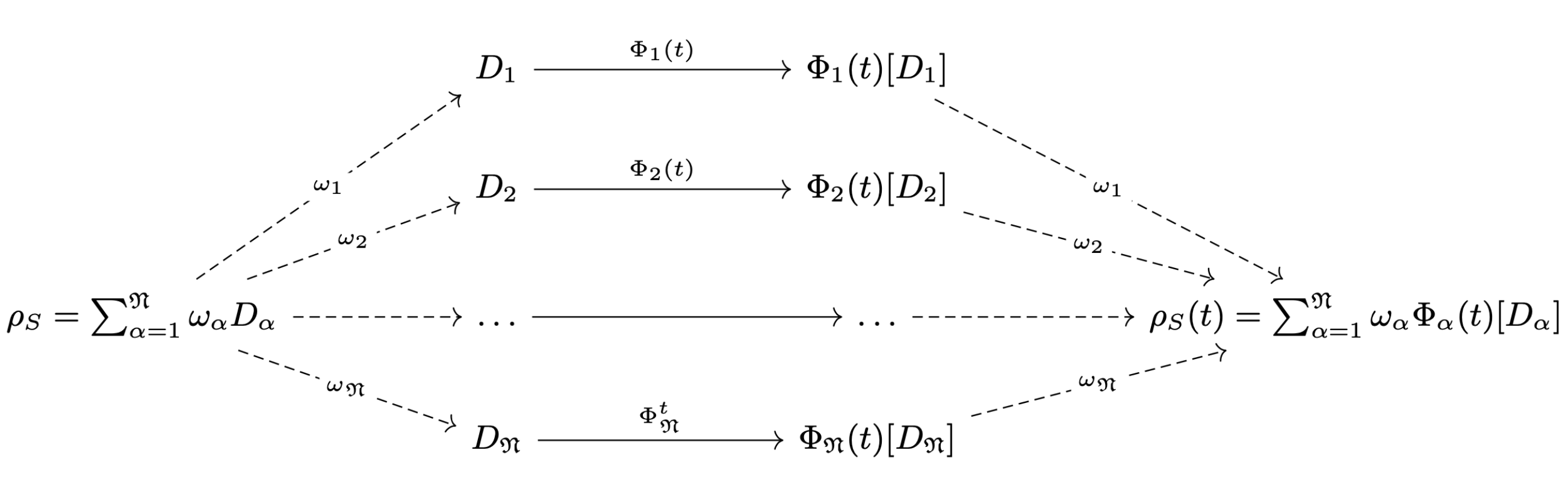}
  \caption{Graphical illustration of the decomposition of the reduced-state evolution in Eq.~(\ref{eq:opdt}). The initial state $\rho_S$ can be written as a combination of operators in the family $\{D_{\alpha}\}$ with positive weights $\{\omega_{\alpha}\}$. A different map acts on each operator of the family. The final state is found recombining each evolved operator with the corresponding initial weight.}\label{fig:1}
\end{figure*}

The central point of interest for the characterization of open-system dynamics is that 
the decomposition in Eq.(\ref{eq:opd}) allows us to express the reduced state at time $t$ via a family of maps that
are CPTP and that are defined on operators
on $\mathcal{H}_S$ only. In fact, replacing Eq.(\ref{eq:opd})
into Eq.(\ref{eq:rst}), one gets
\begin{align}\label{eq:opdt}
	\rho_S(t)&=\sum_{\alpha=1}^{\mathfrak{N}} \omega_{\alpha} \Phi_{\alpha}(t)[D_{\alpha}],
\end{align}
where 
\begin{align}\label{eq:phialpha}
	\Phi_{\alpha}(t): \hs&\to \hs \\
							A &\mapsto \Phi_{\alpha}(t)[A]=\Tr_E[U(t)A\otimes \rho_{\alpha}U(t)^{\dagger}], \notag
\end{align}
so that the $\mathfrak{N}$ CPTP maps $\left\{\Phi_{\alpha}(t)\right\}_{1,\ldots, \mathfrak{N}}$ on $\hs$ associate the initial reduced state
\begin{equation}\label{eq:inits}
\rho_S = \sum_{\alpha=1}^{\mathfrak{N}} \omega_{\alpha} D_{\alpha}
\end{equation}
with the reduced state at time $t$, as illustrated in Fig.~\ref{fig:1}.
Indeed, in case of an initial product state, i.e., $\mathfrak{N}=1$, we recover the usual
description of the reduced dynamics in terms of a single CPTP map \cite{Breuer2002,Rivas2014}.
The price to pay due to the presence of initial correlations is that we will generally need
$\mathfrak{N}>1$ CPTP maps, but this price is (at least, partially) mitigated by the fact that the same family of maps
can be used for different initial states: as shown in \cite{Silva2019}, one can use the same
set $\left\{\Phi_{\alpha}(t)\right\}_{1,\ldots, \mathfrak{N}}$ for all the states connected by
any local operation on S. 

\subsubsection{Standard projection}
Going back to the perturbative expansion of the reduced dynamics
via projection operator techniques, we first replace the decomposition given by Eq.(\ref{eq:opd}) of the initial state $\rho_{SE}$ into Eq.(\ref{eq:stst}),
so that
the linearity of the maps defined in Eq.(\ref{eq:jj}) leads us to 
\begin{align}
\frac{d}{dt}\rho_S(t)=&\sum_{\alpha=1}^{\mathfrak{N}}\omega_{\alpha} \Big(\mathcal{J}^{(1)}_S(t)[D_{\alpha}\otimes \Delta_{\alpha}] + \mathcal{J}^{(2)}_S(t)[D_{\alpha}\otimes \Delta_{\alpha}] 
\Big) \notag\\&+ \mathcal{J}^{(1)}_S(t)[\rho_S(t)\otimes\overline{\rho}_E] + \mathcal{J}^{(2)}_S(t)[\rho_S(t)\otimes\overline{\rho}_E],\label{eq:prepr}
\end{align}
where we have introduced
\begin{equation}\label{eq:exex}
\Delta_\alpha = \rho_\alpha-\overline{\rho}_E,
\end{equation}
i.e., the differences between each environmental statistical operator $\rho_\alpha$ in the decomposition in Eq.(\ref{eq:opd})
and the reference state associated to the standard projection.
Since the maps $\mathcal{J}^{(1,2)}_S(t)$ in Eq.(\ref{eq:prepr}) 
are applied to factorized self-adjoint operators, 
we can exploit the decomposition of the interaction Hamiltonian as
\cite{Breuer2002,Rivas2012}
$H_{I}=\sum_{j}  A_j \otimes B_j$,
with self-adjoint operators $A_j$ and $B_j$,
to express the second order TCL equation in a more explicit form.
In the interaction picture we have 
\begin{equation}
\label{eq:H_int_eigenop_repr}
	H_{I}(t)=\sum_{j}  A_j(t) \otimes B_j(t),
\end{equation}
with $A_j(t) = e^{i H_S t} A_j e^{-i H_S t}$ and $B_j(t) = e^{i H_E t} B_j e^{-i H_E t}$,
so that the corresponding Liouville map reads
$\mathcal{L}(t)[\bullet] =-i\sum_j [A_j(t)\otimes B_j(t), \bullet]$. 
Replacing this expression into Eq.(\ref{eq:jj}),
we encounter the functions
\begin{align}
	\mathfrak{F}_{j_1 j_2}^ {(\rho_\alpha,\overline{\rho}_E)}(t_1,t_2)=&\braket{B_{j_1}(t_1)B_{j_2}(t_2)}_{\rho_\alpha}\notag\\
	&-\braket{B_{j_1}(t_1)}_{\overline{\rho}_E}\braket{B_{j_2}(t_2)}_{\rho_\alpha},\label{eq:notagf} \\
	\mathfrak{G}_{j_2 j_1} ^ {(\rho_\alpha,\overline{\rho}_E)}(t_2,t_1)=&\braket{B_{j_2}(t_2)B_{j_1}(t_1)}_{\rho_\alpha} \notag\\
	&-\braket{B_{j_1}(t_1)}_{\overline{\rho}_E}\braket{B_{j_2}(t_2)}_{\rho_\alpha},\label{eq:notagg}
\end{align}
where we use the common notation
\begin{equation}\label{eq:exex2}
\braket{O}_{\rho} = \Tr \left[O \rho \right];
\end{equation}
importantly, for $\rho_\alpha=\overline{\rho}_E$, the functions in Eqs.(\ref{eq:notagf}) and
(\ref{eq:notagg}) 
reduce to the usual covariance functions of the environmental interaction operators with respect to the  
reference state $\overline{\rho}_E$, that is
\begin{equation}\label{eq:cov}
	\mathfrak{F}_{j_1 j_2}^ {(\overline{\rho}_E,\overline{\rho}_E)}(t_1,t_2)=\mathfrak{G}_{j_1 j_2} ^ {(\overline{\rho}_E,\overline{\rho}_E)}(t_1,t_2)=\mbox{Cov}_{j_1,j_2}^{\overline{\rho}_E}(t_1,t_2). 
\end{equation}

The functions in Eqs.(\ref{eq:notagf}), (\ref{eq:notagg}) and (\ref{eq:cov}) allow us to write 
Eq.(\ref{eq:prepr})
as (see also Eq.(\ref{eq:exex}) and (\ref{eq:exex2}))
\begin{align}
\label{eq:TCL_Delta_S}
\frac{d}{dt}\rho_S(t)&=\sum_{\alpha=1}^{\mathfrak{N}}\omega_{\alpha} \Bigg( -ig\sum_j [A_j(t),D_{\alpha}] \braket{B_j(t)}_{\Delta_{\alpha}} \notag\\&
-g^2\sum_{j_1,j_2}\int_0^t d\tau [A_{j_1}(t),A_{j_2}(\tau)D_{\alpha}] \mathfrak{F}^ {(\Delta_{\alpha},\overline{\rho}_E)}_{j_1 j_2}(t,\tau) \notag\\&
 +g^2 \sum_{j_1,j_2}\int_0^t d\tau
[A_{j_1}(t),D_{\alpha}A_{j_2}(\tau)]  \mathfrak{G}_{j_2 j_1}^{(\Delta_{\alpha},\overline{\rho}_E)}(\tau,t) \Bigg) \notag\\&
-ig\sum_j [A_j(t),\rho_S(t)]\braket{B_j(t)}_{\overline{\rho}_E} \notag\\& 
-g^2\sum_{j_1,j_2}\int_0^t d\tau[A_{j_1}(t),A_{j_2}(\tau)\rho_S(t)] \mbox{Cov}_{j_1,j_2}^{\overline{\rho}_E}(t,\tau) \notag\\&
+g^2 \sum_{j_1,j_2}\int_0^t d\tau
[A_{j_1}(t),\rho_S(t)A_{j_2}(\tau)] \mbox{Cov}_{j_2,j_1}^{\overline{\rho}_E}(\tau,t).
\end{align}
The homogeneous part of the second-order TCL equation 
(the last three lines in Eq.(\ref{eq:TCL_Delta_S})) 
does not depend on the initial-state parameters $\omega_{\alpha}$ and $\rho_{\alpha}$: 
The effects of the initial system-environment correlations on the subsequent reduced dynamics is fully encoded into the inhomogeneous part of the equation (first three lines). 
More precisely, the homogeneous part depends on the environmental covariance functions 
with respect to the environmental reference state $\overline{\rho}_E$, while in the inhomogeneous
part there appear the functions in Eqs.(\ref{eq:notagf})
and (\ref{eq:notagg}), which can be seen as generalizations of the covariance functions
accounting for the initial correlations.
In fact, $\mathfrak{F}_{j_1 j_2}^ {(\rho_\alpha,\overline{\rho}_E)}(t_1,t_2)$
and $\mathfrak{G}_{j_1 j_2}^ {(\rho_\alpha,\overline{\rho}_E)}(t_1,t_2)$ include, 
besides the expectation
values of the environmental operators on $\overline{\rho}_E$, their expectation values
and two-time correlation functions on the environmental states
$\rho_\alpha$.
The access to these functions via the reconstruction of the open-system dynamics 
can be at the basis, for example, of noise-spectroscopy protocols in the presence of initial correlations,
as investigated extensively in \cite{Silva2019}.

\subsubsection{Correlated-state projection}
Also in the case of correlated-state projections
we can exploit the decomposition of the initial state $\rho_{SE}$
as in Eq.(\ref{eq:opd}), along with Eq.(\ref{eq:H_int_eigenop_repr}), to apply the maps in Eq.(\ref{eq:jjcorr}) to factorized self-adjoint operators. 
In analogy with Eq.(\ref{eq:prepr}), the evolution
equations take the form
 \begin{align}
\label{eq:dyn_Delta_corr_proj2}
&\frac{d}{dt}\eta_i(t)=\sum_{\alpha =1}^{\mathfrak{N}} \omega_{\alpha}
\left(\mathcal{J}_i^{(1)}(t)[D_\alpha \otimes \tilde{\Delta}_\alpha] 
+ \mathcal{J}_i^{(2)}(t)[D_\alpha \otimes \tilde{\Delta}_\alpha] \right) \notag\\&
+ \sum_{j}\left(\mathcal{J}_i^{(1)}(t)\left[\eta_{j}(t)\otimes \overline{X}_{j}\right]+ 
\mathcal{J}_i^{(2)}(t)\left[\eta_{j}(t)\otimes \overline{X}_{j}\right]\right),
\end{align}
where we have defined 
\begin{equation}
\tilde{\Delta}_\alpha = \rho_\alpha - \sum_i \overline{X}_i \Tr_E\left[\overline{Y}_i \rho_\alpha\right].
\end{equation}

Thus, we have now a system of coupled
differential equations, as a consequence of the general definition of the projection in 
Eq.(\ref{eq:repr_theor_proj}).
Using the definitions in Eq.(\ref{eq:jjcorr}) one obtains evolution equations for the components 
$\eta_i(t)$ as reported in Appendix \ref{app:a}, in which correlations functions appear that however lack the transparent physical reading in terms of covariance functions obtained 
for a product-state projection. Once we know the evolution for each different component $\eta_i(t)$, 
we can then reconstruct the reduced state at time $t$ as (see Eqs.(\ref{eq:rels})
and (\ref{eq:prhosecorr}))
\begin{equation}\label{eq:rhocorr}
	\rho_S(t)=\sum_i \Tr_E[\overline{X}_i]\eta_i(t).
\end{equation}
Let us stress that this is a general feature of correlated-state projections and 
it is indeed analogous to what happens with the decomposition of the dynamics in Eq.(\ref{eq:opd}),
see Eq.(\ref{eq:opdt}). 

The considered treatments considerably simplify if the projected state $\mathcal{P}[\rho_{SE}(t)]$ is a separable state.
If we restrict to the case where 
the operators $\{\overline{X}_i\}$ and $\{\overline{Y}_i\}$ are positive, $\overline{X}_i \geq 0$ 
and $\overline{Y}_i \geq 0$, and such that $\mbox{Tr}_E[\overline{X}_i]=1$,
$\sum_i \overline{Y}_i = \mathbbm{1}_E$ and 
$\Tr_E[\overline{X}_i \overline{Y}_j]=\delta_{ij}$
\footnote{An example of a family of operators satisfying these conditions is given by \cite{Breuer2007}
$$
\overline{Y}_i = \Pi_i, \quad \overline{X}_i = \frac{\Pi_i \overline{\rho}_E \Pi_i}{\Tr_E[\Pi_i \overline{\rho}_E]}, 
$$
where $\{ \Pi_i\}_i$ is a family of orthogonal projections on $\mathcal{H}_E$
summing up to the identity, and $\overline{\rho}_E$ is a fixed environmental state;
note that since
$\sum_i (\mathbb{1}_S \otimes \Pi_i) \mathcal{P}[\rho_{SE}] (\mathbb{1}_S \otimes \Pi_i) 
=  \mathcal{P}[\rho_{SE}]$, in this case $\mathcal{P}$
actually projects into the set of zero-discord states \cite{Ferraro2010}.},
the conditions in Eq.(\ref{eq:corr_proj_prop_1}) hold, 
and the resulting action of the correlated projection operator in 
Eq.(\ref{eq:prhosecorr})
can be written as
\begin{equation}\label{eq:separ}
 	\mathcal{P}[\rho_{SE}(t)]=\sum_i p_i(t) \rho_{S,i}(t)\otimes \rho_{E,i}.
\end{equation} 
Importantly, the operators $\rho_{E,i}=\overline{X}_i$ are environmental states,
and the coefficients $p_i(t) = \Tr_{SE}\big[(\mathbb{1}_S\otimes \overline{Y}_i)\rho_{SE}(t)]$
are positive and sum up to 1, which means that the projection $\mathcal{P}$
provides us with a representation of the relevant part of the global state 
$\mathcal{P}[\rho_{SE}(t)]$ as in Eq.(\ref{eq:opd}); even more, also the operators $\rho_{S,i}(t)$ defined by
(see Eq.(\ref{eq:etai}))
$p_i(t) \rho_{S,i}(t) =  \Tr_{E}\big[(\mathbb{1}_S\otimes \overline{Y}_i)\rho_{SE}(t)]$
are proper open-system statistical operators, meaning that the relevant part in Eq.(\ref{eq:separ})
actually consists of a separable state. Conversely, whenever the initial global state is a separable
state, $\rho_{SE} = \sum_i p_i \rho_{S,i}\otimes \rho_{E,i}$, and it is possible to introduce
a family of positive environmental operators $\{\overline{Y}_i \geq 0\}$ such that $\sum_i \overline{Y}_i = \mathbb{1}_E$ and $\Tr_E[\rho_{E,i}\overline{Y}_j]=\delta_{ij}$,
choosing the correlated projection operator as in Eq.(\ref{eq:repr_theor_proj})
(with $\overline{X}_i = \rho_{E,i}$)
would remove the inhomogeneity in Eq.(\ref{eq:corr_proj_eta_dyn_TCL}),
since $\mathcal{Q}[\rho_{SE}]=0$, and the representation of 
$\mathcal{P}[\rho_{SE}]=\rho_{SE}$ as in Eq.(\ref{eq:prhosecorr})
would coincide with the representation of $\rho_{SE}$ as in Eq.(\ref{eq:opd}).

\subsection{Adapted projection operator}
\label{sec:adapt-techn-per}
Until now, we have derived a description of the reduced dynamics starting from the TCL equation for the global
unitary evolution with respect to a generic projection $\mathcal{P}$, Eqs.(\ref{eq:rel_part_dyn_TCL})-(\ref{eq:II}),
and, after expanding to the second order the equation for a specific choice of $\mathcal{P}$,
we used the decomposition of the initial global state $\rho_{SE}$ as in Eq.(\ref{eq:opd}) to get an explicit approximated 
master equation for $\rho_S(t)$ 

We will now introduce a different strategy that, instead, takes
the decomposition of $\rho_{SE}$ in Eq.(\ref{eq:opd}) as its starting point. 
Such a decomposition represents any initial global state as a convex combination
of $\mathfrak{N}$ product operators $D_\alpha \otimes \rho_\alpha$, see Eq.(\ref{eq:opd}),
implying that the dynamics of $\rho_S(t)$ can be expressed as the convex combination, 
see Eq.(\ref{eq:opdt}) and Fig.~\ref{fig:1}, 
\begin{equation}\label{eq:datt}
\rho_S(t)=\sum_{\alpha=1}^\mathfrak{N}\omega_{\alpha}D_{\alpha}(t)
=\sum_{\alpha=1}^\mathfrak{N}\omega_{\alpha}\Phi_{\alpha}(t)[D_{\alpha}],
\end{equation}
fixed by the maps $\Phi_{\alpha}(t)$, which in the interaction picture read 
(compare with Eq.(\ref{eq:phialpha}))
\begin{equation}\label{eq:phialphaapt}
	\Phi_{\alpha}(t)[\bullet]=\Tr_E\left[U_I(t)(\bullet\otimes\rho_{\alpha})U_I^{\dagger}(t)\right].
\end{equation}
Our basic idea is now to treat each of these contributions independently, in this way getting
an equation of motion for each component $D_{\alpha}(t)$, rather than for the entire state. 

Hence, for any environmental state $\rho_{\alpha}$, let us introduce a product-state
projection 
\begin{equation}\label{eq:pprodalpha}
\mathcal{P}_{\alpha}[\bullet]=\Tr_E\left[\bullet\right]\otimes \rho_{\alpha}.
\end{equation}
The standard technique associated with product-state projection operators recalled in 
Sec.\ref{sec:stand-corr-proj-pp}, when applied to Eq.(\ref{eq:phialphaapt}), leads us to the exact equation 
(compare with Eqs.(\ref{eq:rel_part_dyn_TCL}))
\begin{equation}
\label{eq:rel_part_dyn_TCL-adapt}
\frac{d}{dt}D_\alpha(t)\otimes \rho_\alpha=\mathcal{K}_{\alpha,TCL}(t)[D_{\alpha}(t)\otimes \rho_\alpha],
\end{equation}
where $\mathcal{K}_{\alpha,TCL}(t)$ is as in Eq.(\ref{eq:tclg})
with $\mathcal{P}_\alpha$ instead of $\mathcal{P}$ and $\Sigma_\alpha(t)$
instead of $\Sigma(t)$ defined accordingly.
Quite remarkably, no inhomogeneous term appears, since 
\begin{equation}\label{eq:adv}
(\mbox{Id}_{SE}-\mathcal{P}_{\alpha})[D_{\alpha}\otimes \rho_\alpha] = 0
\end{equation}
for any $\alpha$ as a direct consequence of the choice of projections in 
Eq.(\ref{eq:pprodalpha}).
As anticipated, we call this choice of projections
APO to stress that it is guided by the initial
global state and, in particular, by its decomposition as in Eq.(\ref{eq:opd}).
Crucially, the open-system dynamics resulting from Eqs.(\ref{eq:datt}) and 
(\ref{eq:rel_part_dyn_TCL-adapt}) is fixed by a system of $\mathfrak{N}$
uncoupled homogeneous equations, where $\mathfrak{N}\leq d^2$
for a $d$-dimensional open system, whatever the dimensionality of the environment and the
correlations in the initial global state. This is in stark contrast with the approaches
described in the previous section. A product-state projection as in Eq.(\ref{eq:pprod})
leads to a single equation that is however homogeneous only in the presence of an initial product state; on the other hand, any correlated-state projection as in Eq.(\ref{eq:repr_theor_proj}) allows for homogeneous equations
for a wider class of initial global states, including separable ones, but it involves a coupled system
of equations, whose number is fixed by the cardinality of the set of indices $\{i\}$, which is generally bounded by the square of the environment dimension.

From Eq.(\ref{eq:rel_part_dyn_TCL-adapt}) it is straightforward
to introduce a perturbative expansion associated with the APO technique. 
Since the latter is defined by a family of product-state projections, see Eq.(\ref{eq:pprodalpha}), we can follow
exactly the same lines that led us from Eq.(\ref{eq:pprod}) to Eq.(\ref{eq:TCL_Delta_S}), but this time
without any inhomogeneous contribution, getting
\begin{align}
\label{eq:ref_TCL_Q}
&\frac{d}{dt}D_{\alpha}(t)=
-i g\sum_j [A_j(t),D_{\alpha}(t)] \braket{B_j(t)}_{\rho_{\alpha}} \\& 
-g^2\sum_{j_1,j_2}\int_0^t d\tau[A_{j_1}(t),A_{j_2}(\tau)D_{\alpha}(t)] \mbox{Cov}_{j_1,j_2}^{\rho_{\alpha}}(t,\tau) \notag\\&
+g^2 \sum_{j_1,j_2}\int_0^t d\tau
[A_{j_1}(t),D_{\alpha}(t)A_{j_2}(\tau)] \mbox{Cov}_{j_2,j_1}^{\rho_{\alpha}}(\tau,t).\notag
\end{align}
The second order expansion of the APO TCL master equation is thus fixed solely by the expectation values
and covariance functions $\mbox{Cov}_{j_1,j_2}^{\rho_\alpha}(t_1,t_2)$ of the environmental operators with respect to the environmental states
$\rho_\alpha$, where $\mbox{Cov}_{j_1,j_2}^{\rho_\alpha}(t_1,t_2)$ is defined as in Eq. \eqref{eq:cov} with $\overline{\rho}_E$ replaced by $\rho_\alpha$.
Comparing Eq.(\ref{eq:ref_TCL_Q}) with Eq.(\ref{eq:TCL_Delta_S}), we can see how, as a consequence of the dependence of the projections $\mathcal{P}_\alpha$
on the environmental
states $\rho_\alpha$,
the APO master equation encloses the full dependence on the initial correlations in a time
homogeneous term, which is essentially what allows one to avoid a time inhomogeneous
contribution for any initial state. Importantly, the APO expansion yields uncoupled homogenous
equations for the operators $\{ D_{\alpha}(t) \}$, at variance with
the case of correlated projections leading to coupled equations for
the $\{ \eta_i(t) \}$ operators.

\section{Examples}
\label{sec:examples}
We consider now two case study, in order to compare the descriptions of the open-system dynamics provided by the perturbative expansions obtained with, respectively, the standard projection operator
technique discussed in Sec.\ref{sec:stand-corr-proj} and the APO
technique introduced in Sec.\ref{sec:adapt-techn-per}.
The first model we take into account, a two-level system undergoing pure decoherence, can be solved exactly \cite{Breuer2002}, 
which also allows us
to compare the two perturbative techniques with the exact solution.
The second model, a damped two-level system in a bosonic bath, is not exactly solvable,
while it includes both
decoherence and dissipation effects induced by the interaction with the environment, thus leading
to a richer open-system dynamics.

\subsection{Exactly solvable dephasing model}
\label{sec:exactly-solv-deph}
Whenever the loss of coherence with respect to the eigenbasis of the free system
Hamiltonian occurs on a much faster time scale than the other effects due to the interaction
with the environment, the pure-dephasing (or pure decoherence) microscopic modeling \cite{Skinner1986,Breuer2002} yields a satisfactory characterization of the open-system dynamics; this is the case in a variety of relevant physical systems, including
quantum-optical \cite{Liu2011,Liu2018} and condensed-matter \cite{Hall2014,Haase2018}
ones.

Thus, let us consider a two-level system, $\mathcal{H}_S=\mathbb{C}^2$,
and its environment such that their global unitary evolution is fixed by a Hamiltonian
as in Eq.(\ref{eq:h}) with
\begin{equation}\label{eq:pdeph}
H_S = \frac\varsigma 2 \sigma_3, \quad H_{I} = \sigma_3 \otimes B,
\end{equation}
where $\sigma_3$ is the $z$-Pauli matrix 
($\sigma_1$ and $\sigma_2$ are the $x$- and $y$-Pauli matrices), $\varsigma$ is the free frequency of the two-level
system and $B$ is a generic self-adjoint operator of the environment.
Since $[H_S \otimes \mathbb{1}_E , H_{I}]=0$ the overall unitary evolution
can be determined exactly and, moving to the interaction picture,
we have
$H_{I}(t)=\sigma_3 \otimes B(t)$,
where
$B(t)=e^{i H_E t} B e^{-i H_E t}$,
and then \begin{align}
\label{eq:int_ev_deph}
U_I(t)&=T_{\leftarrow}\exp\left[-i\int_0^t d\tau  H_{I}(\tau)\right]\notag\\
&=\ket{1}\bra{1} \otimes V(t) + \ket{0}\bra{0} \otimes V^{\dagger}(t),
\end{align} 
with $\ket{1}$ and $\ket{0}$ the eigenstates of $\sigma_3$
with respect to the eigenvalues, respectively, $1$ and $-1$, and the unitary operator $V(t)$ 
acting on $\mathcal{H}_E$ that reads
\begin{equation}
V(t)=T_{\leftarrow}\exp\left[-i\int_0^t d\tau B(\tau)\right].
\end{equation}
Having the explicit expression of the global unitary, we can get the reduced state
at time $t$ for any initial state $\rho_{SE}$, possibly including system-environment 
correlations.
Let
$\rho_{j j}(t) = \bra{j} \rho_S(t) \ket{j}$, $j=0,1$, and 
$\rho_{j k}(t) = \bra{j} \rho_S(t) \ket{k}$, $j\neq k=0, 1$, be the populations
and coherences of the reduced state with respect to the $\sigma_3$ eigenvectors. It is easy to see from Eq.(\ref{eq:int_ev_deph})
that the populations do not change in time, while, introducing the
representation of $\rho_{SE}$ given in Eq.(\ref{eq:opd}), the coherence $\rho_{10}(t)$ at time $t$
can be written as \cite{Silva2019}
\begin{equation}
\label{eq:exact_coher_ev}
	\rho_{10}(t)=\sum_{\alpha=1}^{\mathfrak{N}}\omega_{\alpha} \braket{1|D_{\alpha}|0}\kappa_{\alpha}(t),
\end{equation}
where we defined the generally complex functions
\begin{equation}\label{eq:kappat}
	\kappa_{\alpha}(t)=\Tr_E \left[(V(t))^2\rho_{\alpha}\right];
\end{equation}
of course, $\rho_{01}(t)=\rho^*_{10}(t)$.
Thus, Eqs.(\ref{eq:exact_coher_ev}) and (\ref{eq:kappat}) give us the exact reduced
dynamics, at any time $t$ and for any initial global state $\rho_{SE}$.

In the following, we always consider the decomposition of the initial state $\rho_{SE}$
as in Eq.(\ref{eq:opd})
obtained from the Pauli basis of operators in $\mathcal{L}_2(\mathbbm{C}^2)$.
In this case, the system operators $D_{\alpha}$ are simply given by \cite{Silva2019}
\begin{align}
    D_0 = & \frac{1}{\sqrt{2}}\Big(\mathbb{1}_2-\sigma_1-\sigma_2-\sigma_3\Big), \quad 
D_1 = \frac{1}{\sqrt{2}}\sigma_1,	\notag\\
D_2=& \frac{1}{\sqrt{2}}\sigma_2, \quad
D_3 = \frac{1}{\sqrt{2}}\sigma_3,\label{eq:paulib1}
\end{align}
while the products between the weights $\omega_a$
and the environmental operators $\rho_{\alpha}$ are related to $\rho_{SE}$ by
the positive operators 
\begin{align}
  F_0 =& \frac{1}{\sqrt{2}}\mathbb{1}_2, \quad 
	F_1 = \frac{1}{\sqrt{2}}\Big(\mathbb{1}_2+\sigma_1\Big),	\notag\\
F_2 =& \frac{1}{\sqrt{2}}\Big(\mathbb{1}_2+\sigma_2\Big), \quad
	F_3 = \frac{1}{\sqrt{2}}\Big(\mathbb{1}_2+\sigma_3\Big) \label{eq:paulib2}
\end{align}
via
\begin{equation}
  \omega_\alpha \rho_{\alpha} = \mbox{Tr}_S\left[F_{\alpha} \otimes \mathbbm{1} \rho_{SE}\right].\label{eq:paulib3}
\end{equation}

\subsubsection{Perturbative expansions}
\label{sec:treatm-corr-init}
Moving to the perturbative expansions discussed in Secs.\ref{sec:stand-corr-proj} and \ref{sec:adapt-techn-per}, it can be easily seen that they also yield a description
of the reduced dynamics where the populations do not evolve in time, while
the evolution of the coherence has the same form as in Eq.(\ref{eq:exact_coher_ev}), but with
time-dependent functions that are different from the exact case.

Let us start from the second-order equation (\ref{eq:TCL_Delta_S}) obtained from a standard projection as in Eq.(\ref{eq:pprod}). The interaction Hamiltonian in the interaction picture is as in Eq.(\ref{eq:H_int_eigenop_repr}) with a single term, such that the open-system interaction operator 
$\sigma_3$ does not depend on time; moreover, we have for any operator $O$ acting on $\mathbb{C}^2$
\begin{eqnarray}
	&&\braket{j|[\sigma_3,O] |j}=\braket{j|[\sigma_3,\sigma_3 O] |j}	\label{eq:properties_deph_coher}\\
	&&=\braket{j|[\sigma_3,O\sigma_3] |j}=0 \quad j =0,1;\nonumber\\
	&&\braket{1|[\sigma_3,O] |0}=-\braket{1|[\sigma_3,\sigma_3 O] |0} \nonumber\\
	&&=\braket{1|[\sigma_3,O\sigma_3] |0}=-2\braket{1|O|0}.
\nonumber
\end{eqnarray}
The first relation implies that the populations do not evolve in time, while the second relation
leads us to
\begin{eqnarray}
	\frac{d}{dt}\rho^{TCL}_{10}(t)&=& \sum_{\alpha=1}^{\mathfrak{N}}\omega_{\alpha}
	\braket{1|D_{\alpha}|0}h^{(\Delta_{\alpha},\overline{\rho}_E)}(t) \nonumber\\
	&&+ h^{(\overline{\rho}_E,\overline{\rho}_E)}(t) \rho^{TCL}_{10} (t),\label{eq:forg}
\end{eqnarray}
with
\begin{align}
	h^{(\Delta_{\alpha},\overline{\rho}_E)}(t)&=2ig\braket{B(t)}_{\Delta_{\alpha}} - 4g^2 \int_0^t d\tau \mathfrak{Re}\Big[\mathfrak{F}^{(\Delta_{\alpha},\overline{\rho}_E)}(t,\tau)\Big],\nonumber\\
	h^{(\overline{\rho}_E,\overline{\rho}_E)}(t)&=2ig\braket{B(t)}_{\overline{\rho}_E} - 4g^2 \int_0^t d\tau \mathfrak{Re}\Big[\mbox{Cov}^{(\overline{\rho}_E)}(t,\tau)\Big], \label{eq:deph_h_funct:def}
\end{align}
where recall that $\Delta_\alpha$ is defined as in Eq.(\ref{eq:exex}),
while $\mbox{Cov}^{(\overline{\rho}_E)}(t,\tau)$ is the covariance function of the environmental
interaction operator $B(t)$ on the reference state $\overline{\rho}_E$, see Eq.(\ref{eq:cov}),
and $\mathfrak{F}^{(\Delta_{\alpha},\overline{\rho}_E)}$ is its generalization involving the expectation
values with respect to both $\overline{\rho}_E$ and $\Delta_{\alpha}$, see Eq.(\ref{eq:notagf});
note that we use the label TCL to denote the state obtained via the standard second order
TCL expansion.
The solution of Eq.(\ref{eq:forg}), with initial condition (see Eq.(\ref{eq:inits}))
\begin{equation}\label{eq:missing}
\rho_{10}(0)= \sum_{\alpha=1}^\mathfrak{N}\omega_{\alpha}\braket{1|D_{\alpha}|0}
\end{equation}
can be written as
\begin{equation}\label{eq:tcltcl}
\rho_{10}^{TCL}(t)=\sum_{\alpha=1}^\mathfrak{N}\omega_{\alpha}\braket{1|D_{\alpha}|0} \kappa_{\alpha}^{TCL}(t),
\end{equation}
with
\begin{align}
	\kappa_{\alpha}^{TCL}(t)=&1+ \int_0^t d\tau_1 \exp\left[\int_{\tau_1}^t d\tau_2 h^{(\overline{\rho}_E,\overline{\rho}_E)}(\tau_2)\right]\notag\\ & \qquad\qquad\times h^{(\rho_{\alpha},\overline{\rho}_E)}(\tau_1),\label{eq:kappatcl}
\end{align}
where we used $h^{(\Delta_{\alpha},\overline{\rho}_E)}(t) = h^{(\rho_{\alpha},\overline{\rho}_E)}(t)
- h^{(\overline{\rho}_E,\overline{\rho}_E)}(t)$.

Analogously, the second-order master equation obtained via the APO technique, 
Eq.(\ref{eq:ref_TCL_Q}), can be simplified by means of Eq.(\ref{eq:properties_deph_coher}),
leading to time-independent populations and to
\begin{align}\label{eq:ext35}
	\frac{d}{dt}\braket{1|D_{\alpha} (t)|0}=  h^{(\rho_{\alpha},\rho_{\alpha})}(t) \braket{1|D_{\alpha} (t)|0},
\end{align}
where $h^{(\rho_{\alpha},\rho_{\alpha})}$ is defined as in the second line of Eq.(\ref{eq:deph_h_funct:def}),
but with $\overline{\rho}_E$ replaced by $\rho_{\alpha}$.
The solution of Eq.(\ref{eq:ext35}) reads
\begin{equation}
	\braket{1|D_{\alpha} (t)|0}=\braket{1|D_{\alpha}|0} \exp\left[\int_0^t d\tau h^{(\rho_{\alpha},\rho_{\alpha})}(\tau)  \right],
\end{equation}
so that the coherence of the reduced state $\rho_{10}^{APO}(t)$ as described by the APO
technique is
\begin{equation}\label{eq:apoapo}
\rho_{10}^{APO}(t)=\sum_{\alpha=1}^\mathfrak{N}\omega_{\alpha}\braket{1|D_{\alpha}|0} \kappa_{\alpha}^{APO}(t),
\end{equation}
with
\begin{equation}\label{eq:kappaapo}
\kappa_{\alpha}^{APO}(t)=\exp\left[\int_0^t d\tau h^{(\rho_{\alpha},\rho_{\alpha})}(\tau)  \right].
\end{equation}

Summarizing, both the standard TCL and the APO technique lead us to a solution in the form
as in Eq.(\ref{eq:exact_coher_ev}), see Eqs.(\ref{eq:tcltcl}) and (\ref{eq:apoapo}),
but with the exact functions $\kappa_\alpha(t)$ in Eq.(\ref{eq:kappat})
replaced by $\kappa_{\alpha}^{TCL}(t)$ in Eq.(\ref{eq:kappatcl}) and 
$\kappa_{\alpha}^{APO}(t)$ in Eq.(\ref{eq:kappaapo}).

\subsubsection{Dephasing of polarization degrees of freedom}
\label{sec:deph-polar-degr}
To make an explicit comparison among the exact functions $\kappa_{\alpha}(t)$
and the approximated ones $\kappa_{\alpha}^{TCL}(t)$ and $\kappa_{\alpha}^{APO}(t)$, 
we need to specify the 
environmental interaction operator $B(t)$ and the initial global state $\rho_{SE}$.
Hence, we consider a simple instance of the pure-dephasing model, where the environment
is a single continuous degree of freedom, i.e., $\mathcal{H}_E = L^2(\mathbb{R})$.
This model is associated, for example, with the evolution of 
a photon going through a quartz plate,
which has been extensively studied both theoretically and experimentally within
the context of non-Markovian quantum dynamics \cite{Liu2011,Smirne2011,Cialdi2017,Liu2018}.

Hence, let $B$ be the environmental interaction operator defined as 
\begin{equation}\label{eq:deppol}
	B=\frac{\xi }{2}\int dQ ~Q \ket{Q}\bra{Q}, 
\end{equation}
where $\xi$ is a dimensionless parameter fixing the strength of the system-environment coupling
(we set $g=1$ for the coupling parameter used in the previous sections);
in the case of a photon going through a quartz plate, $\xi$ is the difference between the refractive index in the horizontal and vertical polarization,
while $Q$ is associated with the momentum of the photon, focusing on its propagation in one direction;
note that a formally identical model has been considered in the context of dynamical decoupling, identifying the continuous degree of freedom
with the position of a particle moving in one dimension \cite{Arenz2015}.
From Eq.(\ref{eq:kappat}), it is easy to see that the exact dynamics is fixed by the functions
\begin{equation}\label{eq:2ex}
	\kappa_{\alpha}(t)= \int dQ e^{-i{\xi }Q t} p_{\alpha}(Q),
\end{equation}
where we introduced
\begin{equation}\label{eq:paint}
	p_{\alpha}(Q)=\braket{Q|\rho_{\alpha}|Q},
\end{equation}
i.e., the momentum probability density for the environmental state $\rho_\alpha$; the exact $\kappa_{\alpha}(t)$
is then the corresponding characteristic function.
If we further introduce the first and second moments of the probability $p_{\alpha}(Q)$,
\begin{align}
	m_{\alpha}&=\int dQ~Q~ p_{\alpha}(Q),\nonumber\\
\label{eq:def_2_moment}
	m^{(2)}_{\alpha}&=\int dQ~Q^2 p_{\alpha}(Q),
\end{align}
along with the variance 
\begin{equation}\label{eq:ssigma2}
\sigma^2_{\alpha}=m^{(2)}_{\alpha}-m_{\alpha}^2,
\end{equation}
the second-order TCL expression, see Eq.(\ref{eq:kappatcl}), 
can be written as
\begin{align}
\label{eq:k_alpha_TCL}
	\kappa_{\alpha}^{TCL}(t)=& 1 -e^{-i\xi m_E t} e^{-\frac 12 \xi^2  \sigma^2_E t^2}  \\
	&
	\!\!\!\!\!\!\!\!\!\!\!\!\!\!\!\!
\times	\Bigg(
	i\xi  m_{\alpha}	\int_0^t d\tau e^{i\xi m_E \tau} e^{\frac 12 \xi^2  \sigma^2_E \tau^2}
	 \notag
\\
&		+ {\xi^2 }(m^{(2)}_{\alpha} - m_{\alpha} m_E)\int_0^t d\tau \tau e^{i\xi m_E \tau} e^{\frac 12 \xi^2  \sigma^2_E \tau^2} 
	\Bigg) \notag,
\end{align}
where $m_{E}$ and $\sigma^2_{E}$ are as in 
Eqs.(\ref{eq:def_2_moment}) and (\ref{eq:ssigma2}), but with $\rho_\alpha$ in Eq.\eqref{eq:paint} replaced by the reference
state $\overline{\rho}_E$ used to define the projection operator in Eq.(\ref{eq:pprod}).
In addition, the second-order APO expression, see Eq.(\ref{eq:kappaapo}), is
\begin{equation}
\label{eq:k_alpha_APT}
	\kappa_{\alpha}^{APO}(t)= \exp\left[-i \xi  m_{\alpha} t -\frac 12 \xi^2  \sigma^2_{\alpha}t^2\right].
\end{equation}

We note in particular that the second-order APO technique is equivalent to the 
replacement of the probability distribution $p_{\alpha}(Q)$ in Eq.(\ref{eq:2ex})  
with a Gaussian distribution with the same mean value $m_{\alpha}$ and variance $\sigma_{\alpha}^2$.
Importantly, this guarantees that the second-order APO technique reproduces the exact behavior
in the long-time limit. 
In fact, since $\rho_\alpha$ is a state, due to the Riemann-Lebesgue lemma the Fourier transform
of $p_{\alpha}(Q)$ decays to zero for $t\to+\infty$, so that
\begin{align}
\lim_{t\to+\infty}  \kappa_{\alpha}(t)&= 0.
\end{align}
Indeed, the same is true for $\kappa^{APO}_{\alpha}(t)$, since, as said, it is still defined as the Fourier
transform of a (Gaussian) probability distribution:
\begin{align}\label{eq:infapo}
\lim_{t\to+\infty}  \kappa^{APO}_{\alpha}(t)&= 0.
\end{align}
On the other hand, for the second-order TCL expansion one finds
\begin{align}
\label{eq:lim_k_alpha_TCL}
\lim_{t\to+\infty} \kappa_{\alpha}^{TCL}(t)&= 1-\frac{m^{(2)}_{\alpha}-m_{\alpha}m_E}{m^{(2)}_{E}-m_E^2},
\end{align}
which is generally different from zero (unless the first and second
moments with respect to $\rho_\alpha$ and $\overline{\rho}_E$ coincide).

\subsubsection{Comparison of the expansions}
\label{sec:comp-with-exact}
To proceed further and compare the exact and approximated solutions also in the transient time region, 
we specify a class of initial correlated global states.
We consider pure states of the form
\begin{align}
\ket{\psi} = & C_0 \ket{1}\otimes \int dQ f(Q)\ket{Q} \notag\\
 &+ C_1 \ket{0}\otimes \int dQ f(Q)e^{i\theta(Q)}\ket{Q} \label{eq:pol_freq_pure}
\end{align}
with $|C_1|^2+|C_0|^2=1$ and $\int dQ|f(Q)|^2=1$,
so that there are correlations if and only if the function $\theta(Q)$ is not constant. These states are studied in \cite{Smirne2011,Liu2018} where it is shown how a complete simulation of any qubit dephasing dynamics can be obtained with an appropriate control on $f(Q)$ and $\theta(Q)$, so that indeed they provide an important class of reference states.
For the sake of simplification, we assume $C_0, C_1 \in \mathbbm{R}$ and
\begin{align}
\label{eq:assum_g_even_theta_odd}
	|f(Q)|^2=|f(-Q)|^2,~~\theta(-Q)=-\theta(Q).
\end{align}
Taking into account the Pauli-decomposition introduced in Eqs.(\ref{eq:paulib1})-(\ref{eq:paulib3}),
the environmental-state probabilities in Eq.(\ref{eq:paint}) are 
\begin{align}
p_0(Q)&=p_3(Q)=|f(Q)|^2, \nonumber\\
p_1(Q)&=\frac{1}{\mathcal{N}}|f(Q)|^2\left(1+2C_1 C_0 \cos \theta(Q)\right), \nonumber\\
p_2(Q)&=|f(Q)|^2\left(1-2C_1 C_0 \sin\theta(Q)\right), \nonumber
\end{align}
where $\mathcal{N}$ is a normalization constant warranting
$\int d Q \,p_{1}(Q) =1$ ($p_{2}(Q)$ is already normalized
due to Eq.(\ref{eq:assum_g_even_theta_odd}) and normalization of $f(Q)$).
Using the relations
\begin{eqnarray}
\omega_0 \bra{1}D_0 \ket{0}=\frac{1}{2}(i-1); \,\,\,\,\,\,
&&\omega_1 \bra{1}D_1 \ket{0}=\frac{\mathcal{N}}{2} \label{omegavarie}\\
\omega_2 \bra{1}D_2 \ket{0}=-\frac{i}{2}; \,\,\,\,\,\,
&&\omega_3 \bra{1}D_3 \ket{0}=0, \nonumber
\end{eqnarray} 
one can then show that the exact evolution of the coherence, see Eqs.(\ref{eq:exact_coher_ev}) and (\ref{eq:2ex}), 
can be written as
\begin{equation}
\label{eq:pol_freq_exact_pure}
\rho_{10}(t)=C_1 C_0 \int dQ |f(Q)|^2 e^{i\theta(Q)-i\xi Q t}= C_1 C_0 \kappa(t),
\end{equation}
with
\begin{equation}
	\kappa(t)=\int dQ |f(Q)|^2 e^{i\theta(Q)-i\xi Q t},
\end{equation}
which due to Eq.(\ref{eq:assum_g_even_theta_odd}) is a real function of time.

To determinate the approximated TCL and APO expressions, we need to evaluate the first
and second moments, $m_{\alpha}$ and $m_{\alpha}^{(2)}$ respectively, of the probability distributions $p_{\alpha}(Q)$, see 
Eqs.(\ref{eq:k_alpha_TCL}) and (\ref{eq:k_alpha_APT}). 
The property Eq.~(\ref{eq:assum_g_even_theta_odd}) implies that $p_0(Q)$, $p_1(Q)$ and $p_3(Q)$ are even, so that
$m_0=m_1=m_3=0$; instead, $p_2(Q)$ has an odd contribution
such that $m_2\neq 0$ and, in addition, 
$m_2^{(2)}=m_0^{(2)}=\int dQ ~Q^2 |f(Q)|^2$.
Using these relations and making the choice
$\overline{\rho}_E=\rho_E=\Tr_E [\ket{\psi}\bra{\psi}]$ one determines $\rho_{10}^{TCL}(t)$ and $\rho_{10}^{APO}(t)$ according to Eq.(\ref{eq:k_alpha_TCL}) and Eq.(\ref{eq:k_alpha_APT}) respectively. Further using Eq.(\ref{eq:tcltcl}) and Eq.(\ref{eq:apoapo}) together with Eq.(\ref{omegavarie}) the expression for $\rho_{10}^{TCL}(t)$ and $\rho_{10}^{APO}(t)$ are readily obtained as
\begin{eqnarray}
\label{eq:pol_freq_TCL_ASSUM}
\rho_{10}^{TCL}(t) &=& C_1 C_0 \left(1-\frac{m_1^{(2)}}{m_0^{(2)}} \left(1-e^{-\frac 12 {\xi^2 m_0^{(2)} t^2}}\right) \right) \nonumber \\
&&+\frac 12  \left( 1-\frac{m_1^{(2)}}{m_0^{(2)}} \right) \left(1-e^{-\frac 12 \xi^2 m_0^{(2)} t^2}\right) \nonumber \\
&&-\frac 12 m_2 \xi    e^{-\frac 12 \xi^2m_0^{(2)} t^2}\int_0^t d\tau e^{\frac 12 \xi^2 m_0^{(2)} \tau^2}, 
\end{eqnarray}
and
\begin{eqnarray}
\rho_{10}^{AP0}(t)&=&\frac{i-1}{2}e^{-\frac12 {\xi^2 m_0^{(2)} t^2}} +\frac{\mathcal{N}}{2}e^{-\frac 12 {\xi^2 m_1^{(2)} t^2}} \nonumber\\
&&-\frac{i}{2}e^{-i \xi m_2 t-\frac 12 {\xi^2 \left(m_2^{(2)}-m_2^2\right) t^2}}.
\label{eq:pol_freq_APT_ASSUM}
\end{eqnarray}
We observe that, contrary to the exact solution and the second order TCL approximation,
the second order APO solution presents a non-trivial evolution for the imaginary part of the coherence.\\

\begin{figure}[ht!]
	\centering
		\includegraphics[width=0.85\linewidth]{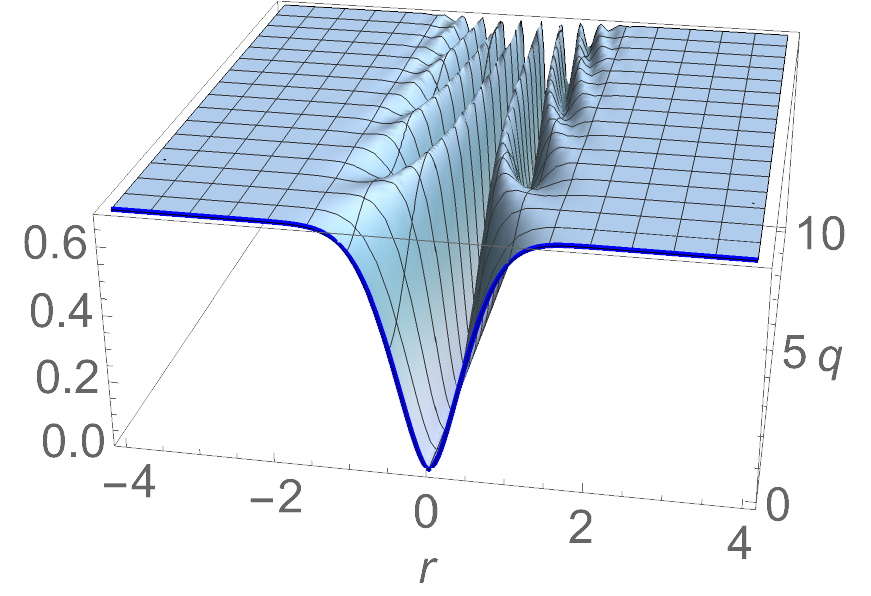}
	\caption{Entropy of entanglement, see Eq.(\ref{eq:entent}), for a system-environment correlated state
	as in Eq.(\ref{eq:pol_freq_pure}) with momentum distribution $f(Q)$ and phase $\theta(Q)$ as in Eq.(\ref{eq:queste2}),
	as a function of $r$ and $q$; the section for $q=0$ (blue, thick line) corresponds to the case 
	where $f(Q)$ is a single
	Gaussian peak, i.e., Eq.(\ref{eq:queste}).}
	\label{fig:extra}
\end{figure}

We now consider specific choices of the functions fixing
the initial global state in Eq.(\ref{eq:pol_freq_pure}). Let us first consider
a symmetric Gaussian $f(Q)$ centered in $Q=0$
and a linear phase $\theta(Q)$, i.e.,
\begin{align}
|f(Q)|^2&=\frac{1}{\sqrt{2\pi\sigma^2}}e^{-\frac{Q^2}{2\sigma^2}}, \quad
\theta(Q)= r \frac{Q}{\sigma}. \label{eq:queste}
\end{align}
For greater $|r|$ the initial reduced state is more mixed, i.e., the pure
state $\ket{\psi}$ is more entangled. Thus, $|r|$ provides an indication on the amount of correlations for this class of pure states; for $r=0$ we have an initial product state, while $|r|\to\infty$ leads to a maximally entangled state.
More in detail, in Fig.\ref{fig:extra} (thick blue line) we show the amount of entanglement for an initial global
state fixed by Eqs.(\ref{eq:pol_freq_pure}) and (\ref{eq:queste}) as a function of $r$, where the entanglement
is quantified by the entropy of entanglement \cite{Horodecki2009}, which is the von Neumann entropy $S$ of the reduced state $\rho_S$, i.e.,
\begin{equation}\label{eq:entent}
   S\left(\mbox{Tr}_E\left[\ket{\psi}\bra{\psi}\right]\right) = -\mbox{Tr}_S\left[\rho_S \log \rho_S\right]. 
\end{equation}
The entropy of entanglement is even with respect to $r$ and it increases monotonically as a function
of $|r|$, already approximating quite closely (up to $1.5{\%}$) the maximum value $1/\sqrt{2}$ for $|r|=2$.

\begin{figure*}
\hspace{0.17\columnwidth}{\footnotesize{\bf (a) $r = -2$}} \hspace{0.23\columnwidth} {
\bf (b) \footnotesize{$r = -1 \quad$}} \hspace{0.27\columnwidth} {\bf (c) \footnotesize{$r = 1 \,\,\,$}} 
\hspace{0.27\columnwidth} {\bf (d) \footnotesize{$r = 2 \qquad$}}
\includegraphics[width=0.5\columnwidth]{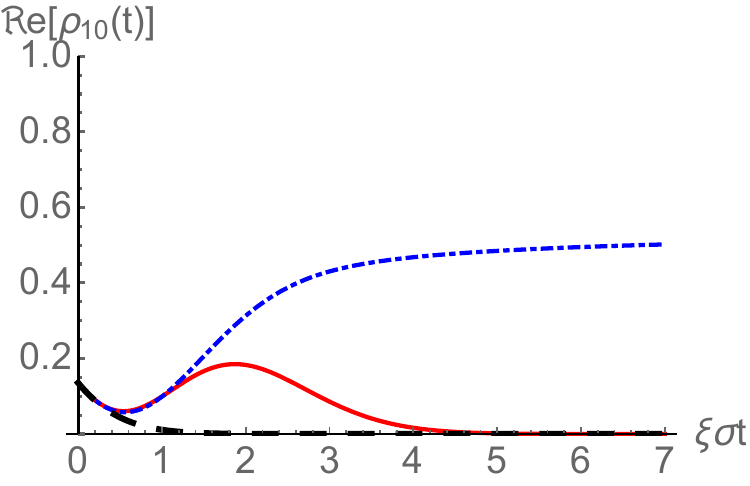}\includegraphics[width=0.5\columnwidth]{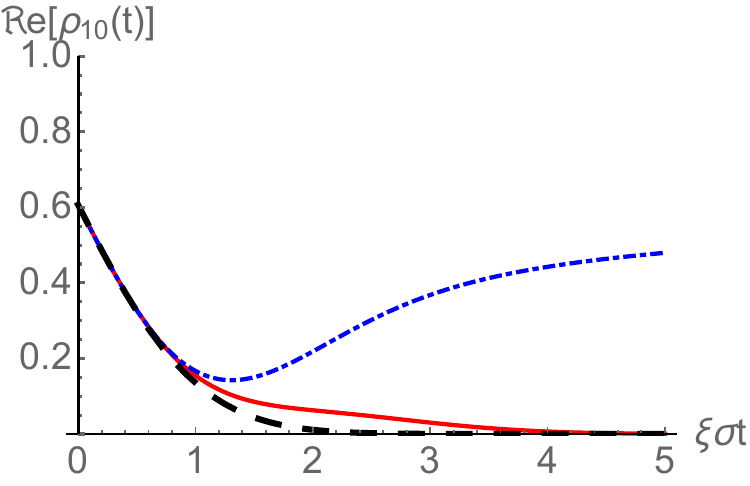}\includegraphics[width=0.5\columnwidth]{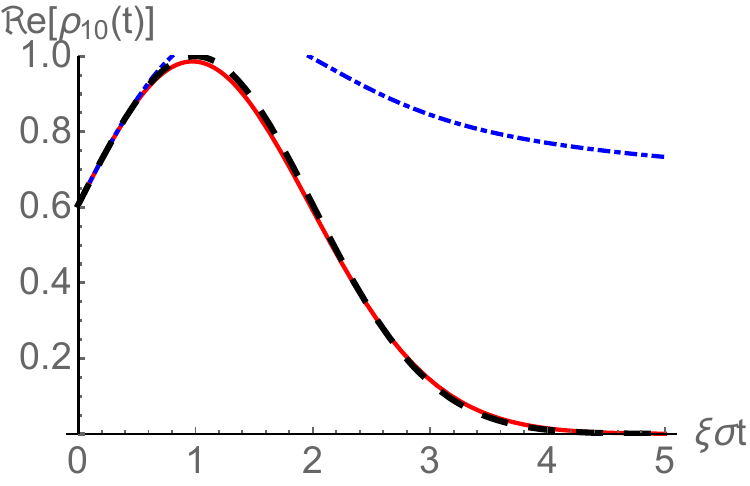}\includegraphics[width=0.5\columnwidth]{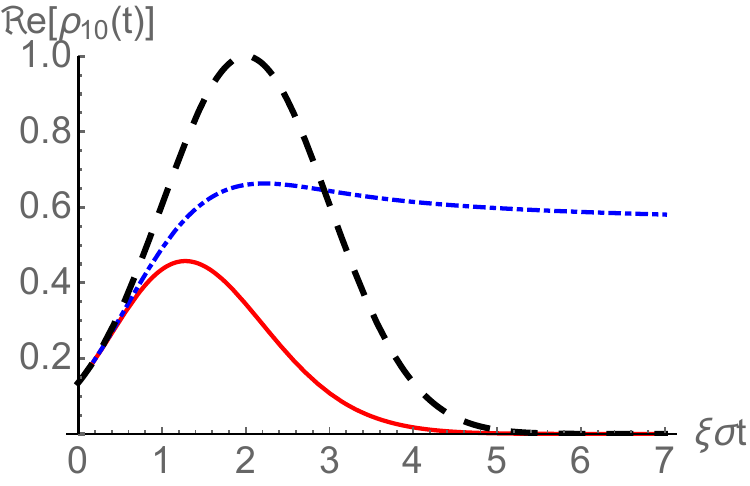}
\caption{{Comparison between exact evolution (black dashed), TCL solution (blue dot-dashed) and APO solution (red solid)
of the real part of the coherence, $\mathfrak{Re}[\rho_{10}(t)]$, for different values of the correlation parameter $r$ in the pure-dephasing dynamics fixed by Eqs.(\ref{eq:pdeph})
and (\ref{eq:deppol}), and initial state given by Eq.(\ref{eq:pol_freq_pure}),
for $f(Q)$ and $\theta(Q)$ as in Eq.(\ref{eq:queste}); 
the values of the parameters are
$C_0= C_1=\frac{1}{\sqrt{2}}$ and $\varsigma=0$;
note that the latter can be read as an indication of a strong system-environment coupling regime
of the pure dephasing.}}
\label{fig:2}
\end{figure*}

In Fig.~\ref{fig:2}, we compare the exact (black dashed line), the TCL (blue dot dashed) and the APO (red solid) solutions 
of the real component of the coherence, $\mathfrak{Re}[\rho_{10}(t)]$, for different values of $r$. We observe that both approximations are in good agreement with the exact solution at short times.
On the other hand, the TCL description departs significantly from the exact solution, 
possibly even becoming unphysical,
at intermediate and long times, while the APO solution is always bounded between $0$ and $1$
and reproduces to a good extent the exact solution during the whole time evolution, for 
small and intermediate values 
of the correlation parameter $|r|$, i.e., for $|r| \lessapprox 1$, and it anyway
captures both the short- and long-time dynamics even for stronger correlations.

\begin{figure*}[t!]
	\centering
	{\bf (a)$\,\,$}\hskip5cm{\bf (b)}\hskip5cm{\bf (c)}\\
	\vspace{0.15cm}
	\includegraphics[width=0.31\linewidth]{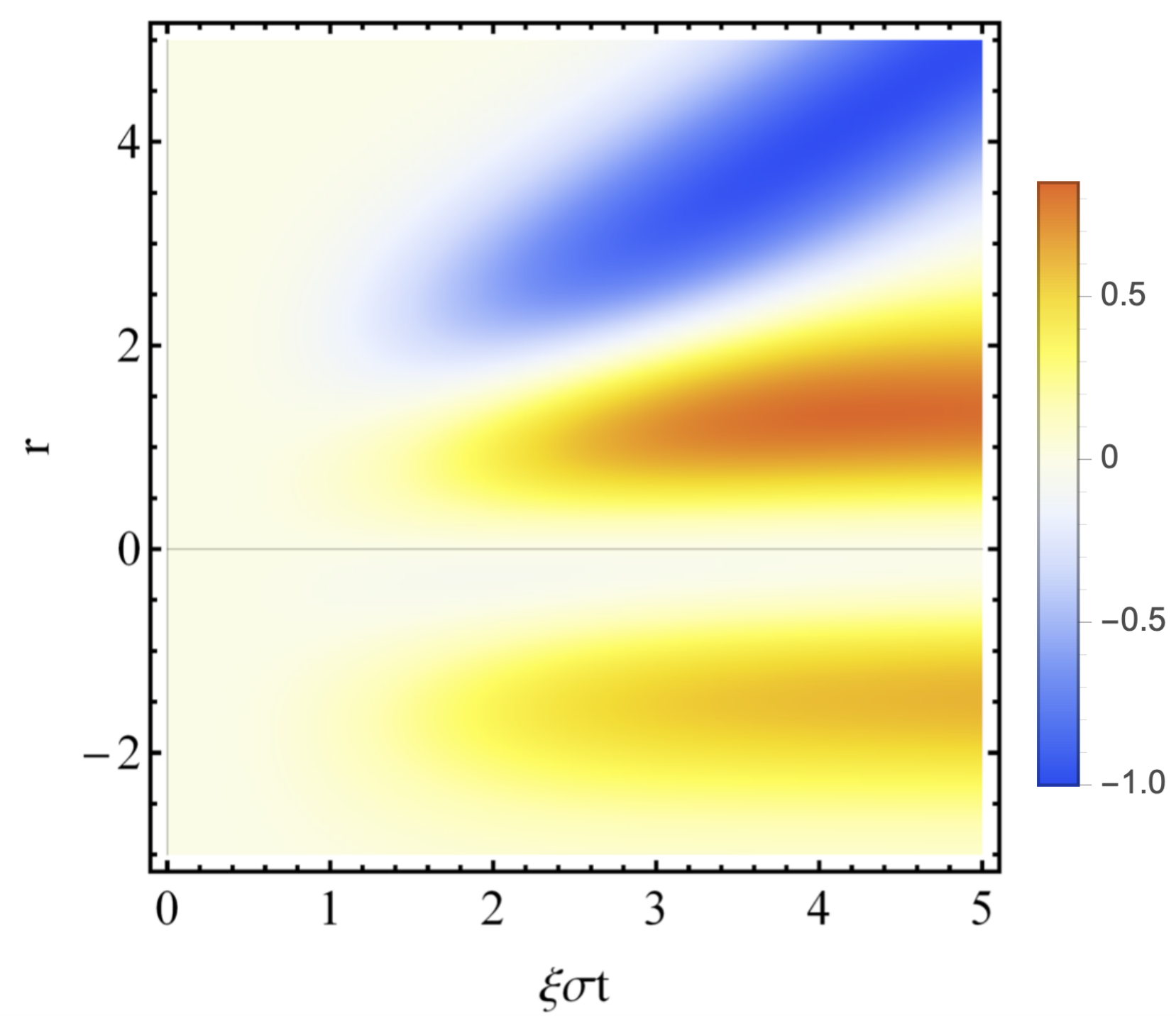}
	\includegraphics[width=0.32\linewidth]{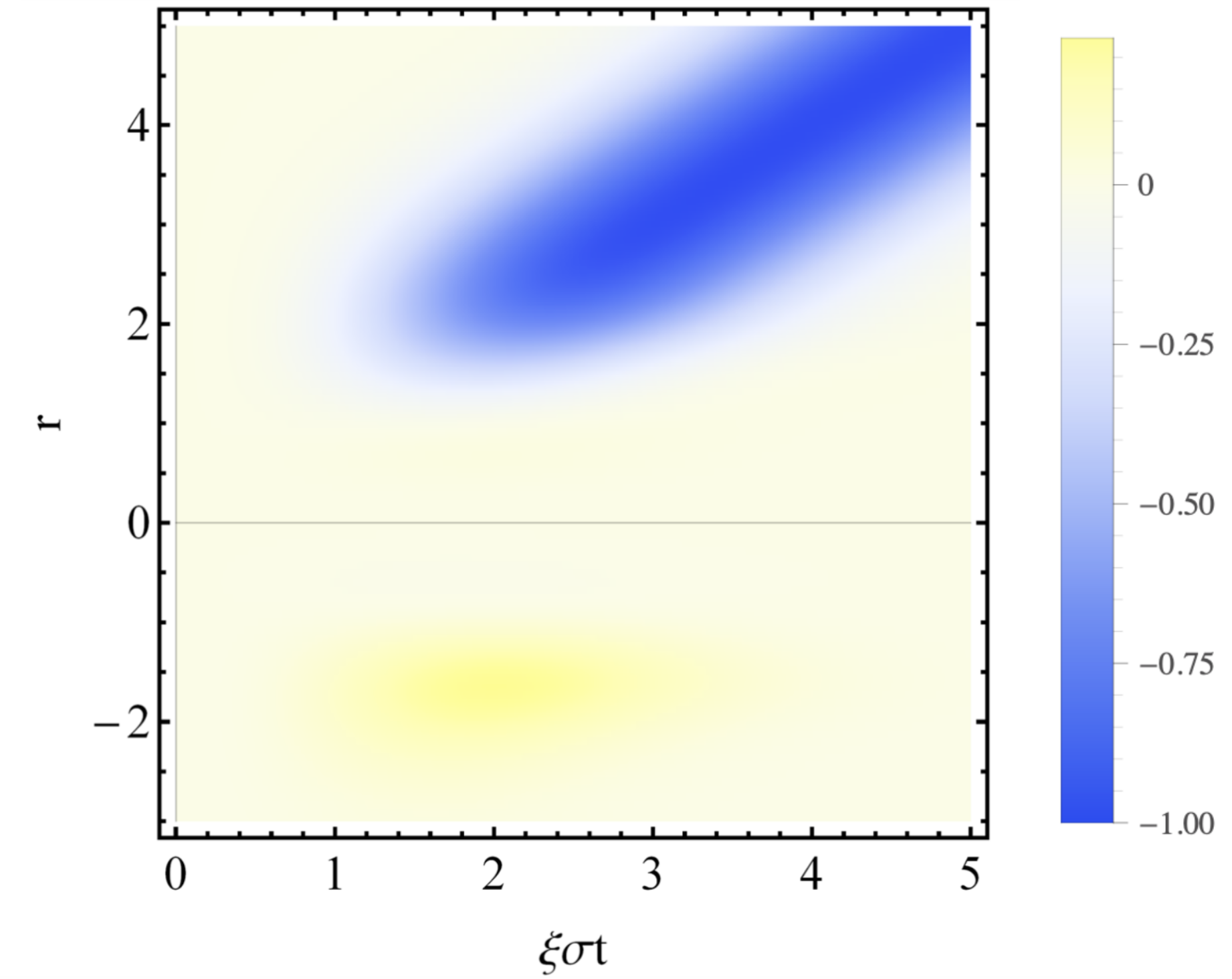}
         \includegraphics[width=0.31\linewidth]{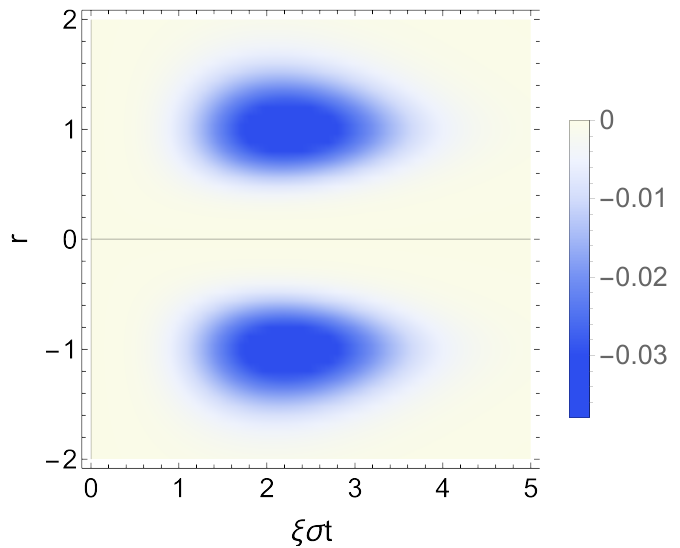}
	\caption{{\bf (a)} Difference between the TCL and exact solution for $\mathfrak{Re}[\rho_{10}(t)]$,
	 difference between the APO and exact solution for {\bf (b)} $\mathfrak{Re}[\rho_{10}(t)]$ 
	 and {\bf (c)} $\mathfrak{Im}[\rho_{10}(t)]$, as a function of time and of the correlation parameter $r$. 
	 The values of the other parameters are as in Fig.~\ref{fig:2}, so that $f(Q)$ and $\theta(Q)$ are as in Eq.(\ref{eq:queste}). The comparison between the panels {\bf (a)} and {\bf (b)} shows the better performance of the APO approximation when compared with the standard TCL.}
	\label{fig:3}
\end{figure*}

The overall better agreement between the second order APO and the exact solution for
$\mathfrak{Re}[\rho_{10}(t)]$ is further confirmed by Fig.~\ref{fig:3} {\bf (a)} and {\bf (b)}. 
There, we consider the difference between the approximated predictions and the exact solution,
as a function both of time and of the correlation parameter $r$.
We notice in both cases the presence of a blue region, associated to a negative error, around 
$\xi\sigma t=r$ for $r>2$. 
This is due to the fact that for large $r$ we have $m_{\alpha}\approx 0$ and $m_{\alpha}^{(2)}\approx \sigma$, so that
$\mathfrak{Re}[\rho^{TCL}_{10}(t)]\approx \mathfrak{Re}[\rho^{APO}_{10}(t)]\approx 0$,
while the exact solution presents a Gaussian peak at $\xi\sigma t=r$.
The horizontal orange regions in the plot referred to the TCL solution is due to the fact that 
the TCL solution converges at long times to a value significantly different from zero; 
in fact, it can be shown that 
$\lim_{t\to+\infty}\mathfrak{Re}[\rho^{TCL}_{10}(t)]=(r^2/2)e^{-\frac{r^2}{2}}$. 
Instead, the APO solution always reproduces the exact behavior at long times, 
see Eq.(\ref{eq:infapo}), which also brings along a better approximation in the transient time
region. The APO solution fits particularly well the exact evolution at all times for $|r|<1$,
while for $|r|>1$ the approximation fails at times $\sigma\xi  t\approx r$, and this is again due to the Gaussian peak of the exact solution.
Finally, in Fig.~\ref{fig:3}~{\bf (c)} we show the evolution of the imaginary part of the coherence in the second-order APO approximation; the deviation from the exact solution (that is always identically equal to zero) is anyway two orders of magnitude smaller than the value of the real part.

\begin{figure*}[t!]
	\centering
		{\bf (a)$\,\,$}\hskip5cm{\bf (b)}\hskip5cm{\bf (c)}\\
	\vspace{0.15cm}
		\includegraphics[width=0.32\linewidth]{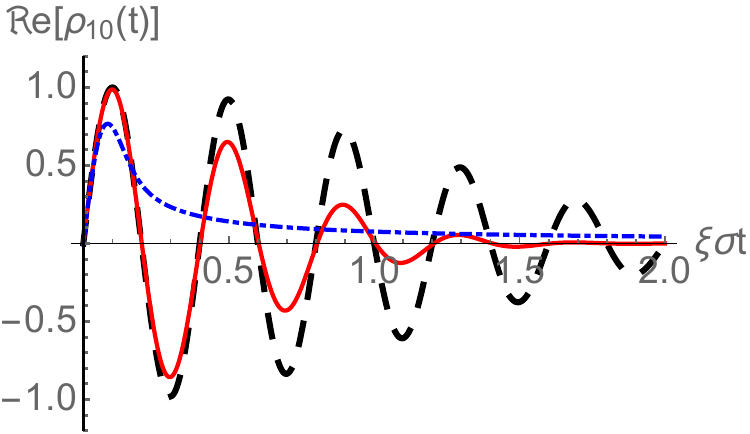}
		\includegraphics[width=0.32\linewidth]{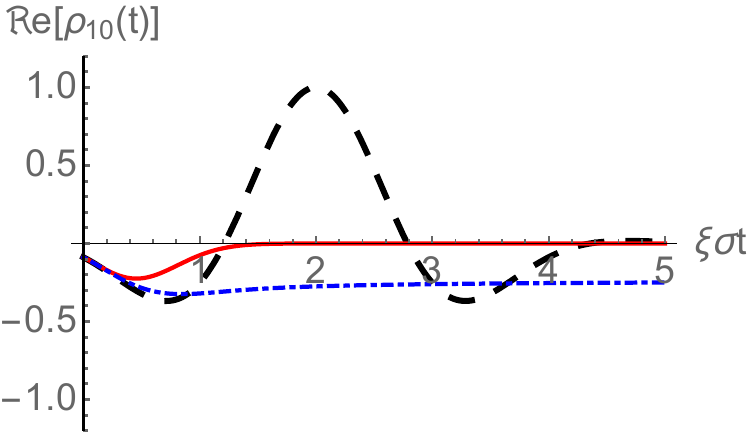}
		\includegraphics[width=0.32\linewidth]{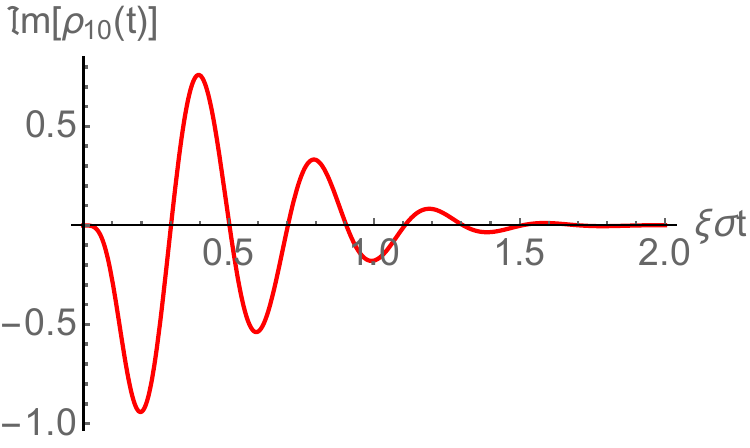}
	\caption{Comparison of the exact evolution (black dashed line), TCL solution (blue dot-dashed) and APO solution (red solid) for {\bf (a)} $r=0.1$ and $q=\pi/2r$  and {\bf (b)} $r=q=2$, 
	of $\mathfrak{Re}[\rho_{10}(t)]$ in the pure-dephasing dynamics fixed by Eqs.(\ref{eq:pdeph})
and (\ref{eq:deppol}), and initial state given by Eq.(\ref{eq:pol_freq_pure}),
for $f(Q)$ and $\theta(Q)$ as in Eq.(\ref{eq:queste2}); 
{\bf (c)} APO solution of $\mathfrak{Im}[\rho_{10}(t)]$ (the exact and TCL values are identically equal 
to 0 at every time) for $r=0.1$ and $q=\pi/2r$; the values of the other parameters are as in Fig.~\ref{fig:2}}
	\label{fig:4}
\end{figure*}

\begin{figure}[t!]
	\centering
	{\bf $\quad$ (a)}\hskip4.1cm{\bf (b)}\\
	\vspace{0.15cm}
		\includegraphics[width=0.48\linewidth]{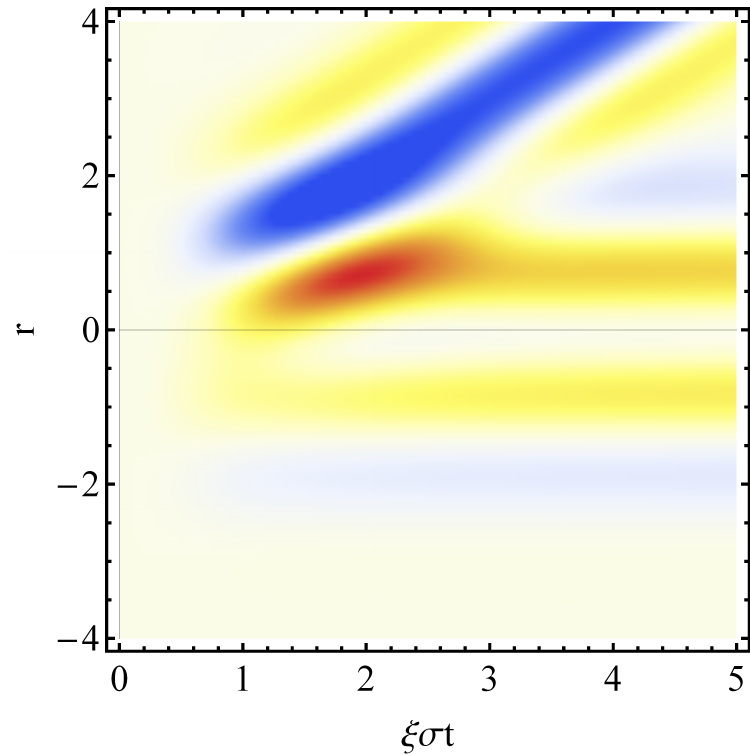}
		\includegraphics[width=0.48\linewidth]{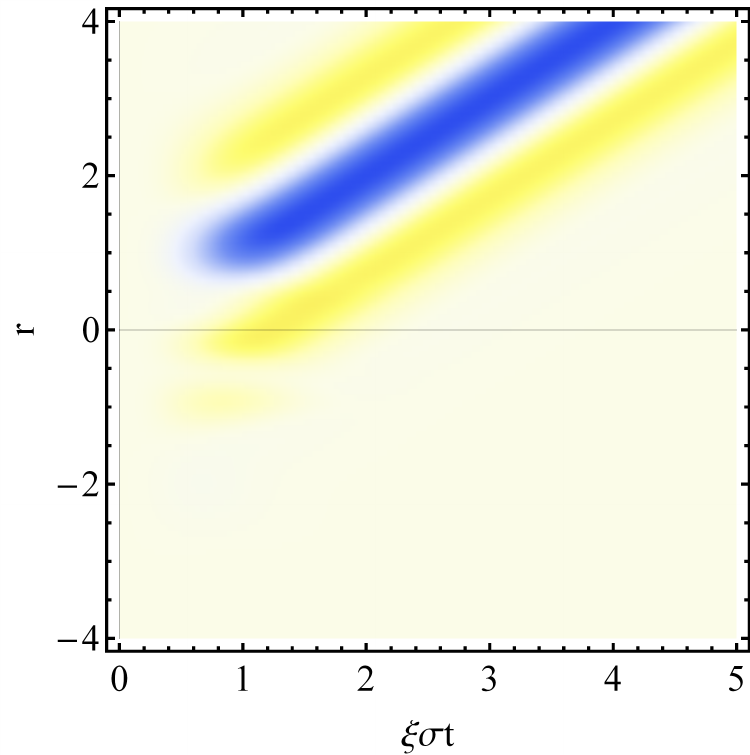}\\
		\vspace{0.15cm}
	{\bf $\quad$ (c)}\hskip4.1cm{\bf (d)}\\
	\vspace{0.15cm}
		\includegraphics[width=0.48\linewidth]{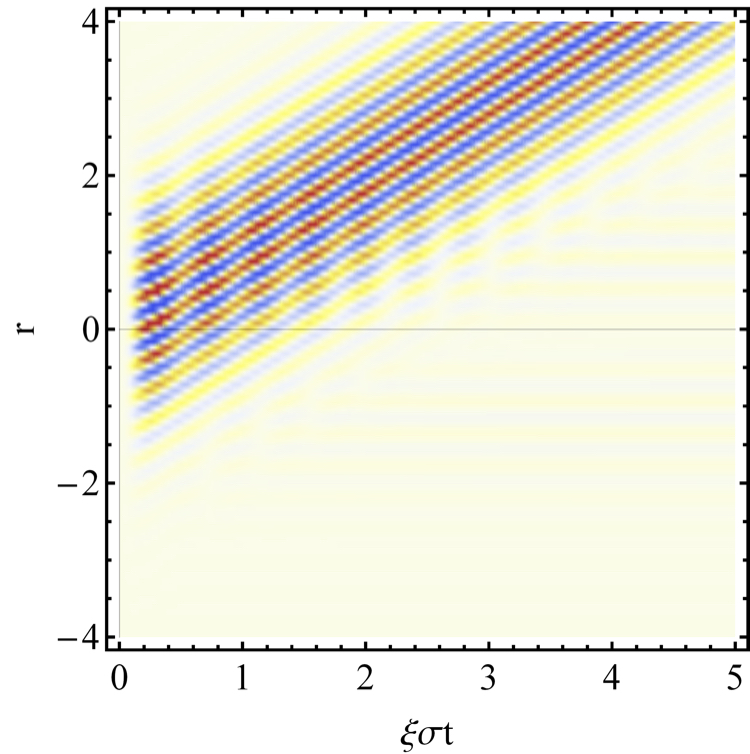}
		\includegraphics[width=0.48\linewidth]{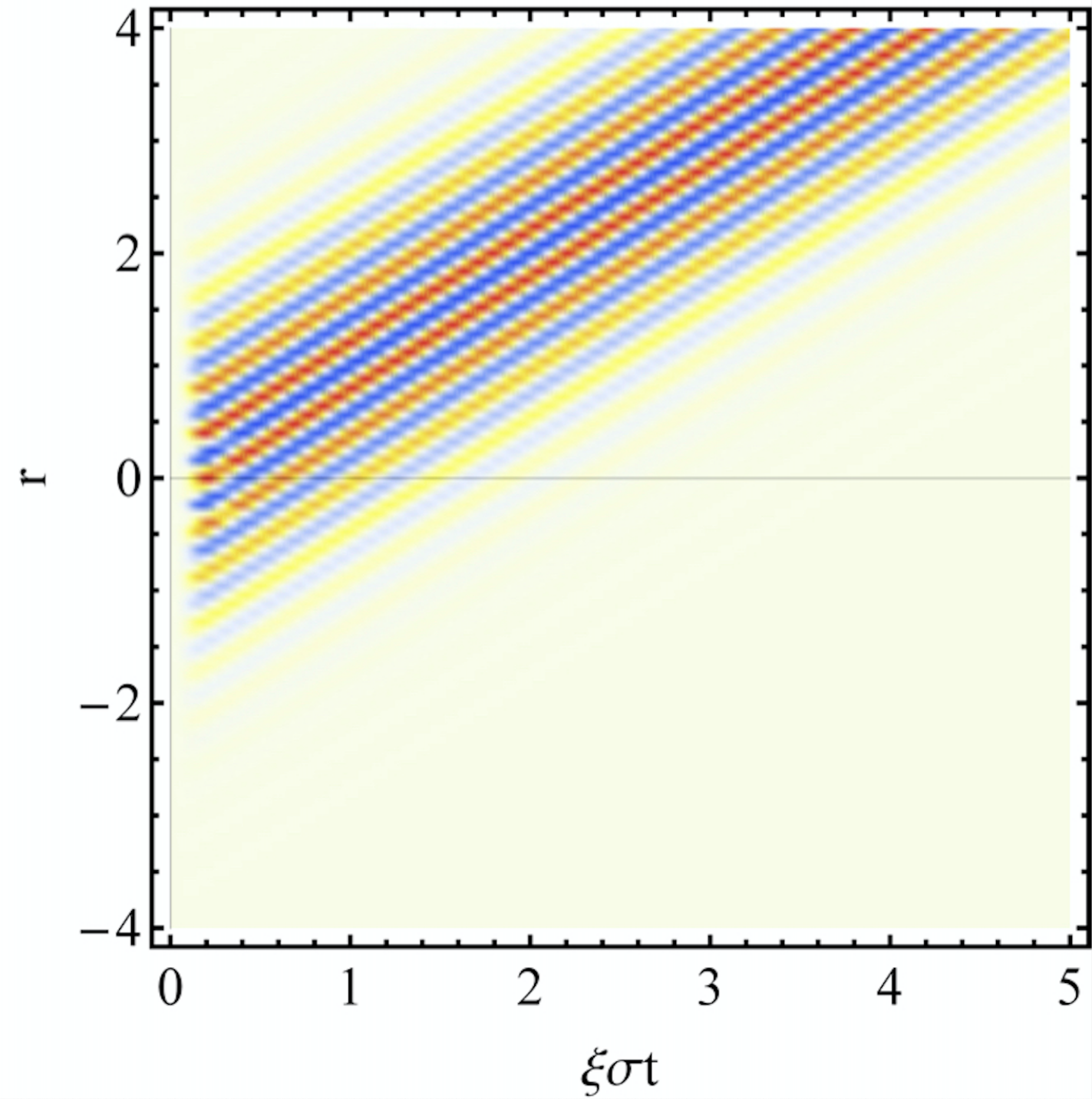}
		\includegraphics[width=0.85\linewidth]{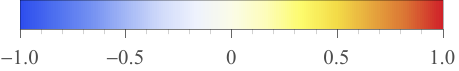}
	\caption{Difference between the TCL and exact solution {\bf (a}, {\bf c)} and
	 difference between the APO and exact solution {\bf (b}, {\bf d)} for $\mathfrak{Re}[\rho_{10}(t)]$, as a function of time and of the correlation parameter $r$ for $q=2$ {\bf (a}, {\bf b)}
	 	 and $q=15$ {\bf (c}, {\bf d)}; the values of the other parameters are as in Fig.~\ref{fig:4}.}
	\label{fig:5}
\end{figure}

In Figs.~\ref{fig:4} and \ref{fig:5}, we consider instead an initial state as in
Eq.(\ref{eq:pol_freq_pure}), but where now the momentum distribution $|f(Q)|^2$ is 
given by the balanced mixture of two symmetric Gaussians centered around $\pm Q_0$:
\begin{align}
|f(Q)|^2&=\frac{1}{2\sqrt{2\pi\sigma^2}}\left(e^{-\frac{(Q-Q_0)^2}{2\sigma^2}} + e^{-\frac{(Q+Q_0)^2}{2\sigma^2}}\right); \nonumber\\
\theta(Q)&= r \frac{Q}{\sigma};\label{eq:queste2}
\end{align} 
indeed, $|f(Q)|^2$ is an even function, so that $\kappa(t)$ is still real.
If we define the ratio $q=Q_0/\sigma$ (for $q\to 0$ the distribution reduces to a
single Gaussian centered in $Q=0$, which is the case of Eq.(\ref{eq:queste})), now the correlations are parametrized by the couple $(r,q)$.
In particular, we observe in Fig.\ref{fig:extra} the entropy of entanglement defined in Eq.(\ref{eq:entent}) as a function
of $r$ and $q$: 
the initial state is maximally entangled for $q r=(2k+1)\pi/2$, $k\in\mathbb{Z}$, which explains
the oscillating behavior as a function of $r$ for values of $q$ different from zero; in addition the maximum
value $1/\sqrt{2}$ is reached for $|r|\rightarrow \infty$ and approximated very closely for $r \gtrsim 2$.

In Fig. \ref{fig:4}, we notice that the exact evolution of $\mathfrak{Re}[\rho_{10}(t)]$ 
presents an oscillation of frequency $q\sigma \xi  /2\pi$,
which is correctly reproduced only by the APO solution, for small values of $r$, while
the TCL solution completely misses such an oscillation.
At higher values of $r$, both the APO and the TCL solutions depart significantly from the exact one
at intermediate times, but the former is indeed still able to properly reproduce the long-time decay.
On the other hand, the APO solution introduces an imaginary component of the coherence 
$\mathfrak{Im}[\rho_{10}(t)]$ (the exact and the TCL solutions are identically equal to 0), which can now take on significant values (of the same order as $\mathfrak{Re}[\rho_{10}(t)]$).
Once again, the overall better agreement between the predictions of the APO description
of $\mathfrak{Re}[\rho_{10}(t)]$ and the exact solution seem to be robust for different values
of the correlation parameters, $r$ and $q$, as shown in Fig.~\ref{fig:5}.
Here, we plot the difference between the approximated solutions and the exact one as functions
of $t$ and $r$, for different values of $q$.
The diagonal stripes that can be observed in both cases are a consequence of the Gaussian peak of the exact solution, which now modulates an oscillation becoming faster for greater $q$ and not captured by the TCL nor the APO solution for high values of $q$. On the other hand, the APO description matches better the exact solution for smaller values of $q$, and especially if one further has small or intermediate values of $r$; here,
the plot of the TCL solution presents also horizontal stripes, in correspondence with a non-zero long-time limit, whose value oscillates from negative values (blue stripes) to positive ones (orange stripes) for different $r$.

\subsection{Damped two-level system in a bosonic bath}
\label{sec:damped-two-level}
In the second model we consider, the open system is still a two-level system,
$\mathcal{H}_S=\mathbbm{C}^2$, which is now interacting with a bosonic environment
exchanging also excitations with it. In particular, we consider a Jaynes-Cummings
form of the interaction Hamiltonian, so that the global Hamiltonian is as in Eq.(\ref{eq:h})
($g=1$)
with
\begin{align}
	H_S&= \frac{\varsigma}{2}\sigma_3,
\quad	H_E = \sum_k \omega_k b^{\dagger}_kb_k, \nonumber\\
	H_{I}&=\sum_k g_k \sigma_+\otimes b_k + g_k^*\sigma_-\otimes b^{\dagger}_k,\label{eq:hdamp}
\end{align}
where $\sigma_{+} = \ket{1}\bra{0}$ and $\sigma_{-} = \ket{0}\bra{1}$
are the raising and lowering operators of the two-level system, with $b_k$ and $b^\dag_k$ the annihilation and creation operators of the $k$-th bosonic mode,
while $g_k$ is its coupling strength with the system.
The interaction picture Hamiltonian can thus be written as in Eq.(\ref{eq:H_int_eigenop_repr}) 
with (having assigned $j \mapsto (k,\pm)$)
\begin{align}
	A_{k,+}(t)&=e^{i\varsigma t}\sigma_+, \quad
	A_{k,-}(t)=e^{-i\varsigma t}\sigma_-,\nonumber\\
	B_{k,+}(t)&=g_k e^{-i\omega_k t}b_k , \quad
	B_{k,-}(t)=g_k^* e^{i\omega_k t}b^{\dagger}_k. \label{eq:hitex2}
\end{align}
Unless one restricts to a single-bath mode \cite{Smirne2010} or to a zero-temperature bath \cite{Garraway1997,Vacchini2010}, this model cannot be solved analytically; moreover, standard projective 
approaches have been applied to it \cite{Breuer2002,Smirne2010} only in the absence of initial correlations.
We will now instead apply both the standard projection technique discussed in Sec.\ref{sec:stand-corr-proj} and the APO technique introduced in \ref{sec:adapt-techn-per} taking into account the presence
of initial correlations.

\subsubsection{Standard and adapted projection second-order master equations}
\label{sec:treatm-corr-init2}
For the sake of simplicity, we focus on initial global states $\rho_{SE}$ such that
the environmental states defining its decomposition as in Eq.(\ref{eq:opd}) satisfy
	\begin{eqnarray}
		\braket{b_{k}}_{\rho_\alpha} = \braket{b_{k} b_{k'}}_{\rho_\alpha} &=&0, \,\,\,
		\braket{b^{\dag}_{k}b_{k'}}_{\rho_\alpha} = \delta_{k, k'} n^\alpha_k, \label{eq:simex2}
	\end{eqnarray}
where we introduced the expectation value of the number operator of the $k$-th mode on $\rho_\alpha$,
$n^\alpha_k = \braket{b^{\dag}_{k}b_{k}}_{\rho_\alpha}$; these conditions generalize
the analogous ones for a thermal state, but, indeed, choosing different $n^\alpha_k$ for different
$\alpha$ allows us to describe initially correlated states.
Moreover, we perform the continuum limit of the bath modes \cite{Breuer2002}
with the replacements $\omega_k \mapsto \omega$, where $\omega$ can take any real positive value,
$\sum_k \mapsto \int_0^\infty d \omega$ and defining the spectral density
\begin{equation}
J(\omega) = \sum_k |g_k|^2 \delta(\omega - \omega_k).
\end{equation}

For the standard projection technique, 
we set 
$\overline{\rho}_E = \rho_E = \sum_{\alpha=1}^\mathfrak{N} \omega_\alpha \Tr[D_\alpha] \rho_\alpha$ in the definition of the projection operator in Eq.(\ref{eq:pprod}),
so that the conditions in Eq.(\ref{eq:simex2}) directly imply similar conditions with respect to  
$\overline{\rho}_E$:
	\begin{eqnarray}
		\braket{b_{k}}_{\overline{\rho}_E} = \braket{b_{k} b_{k'}}_{\overline{\rho}_E} &=&0; \,\,\,
		\braket{b^{\dag}_{k}b_{k'}}_{\overline{\rho}_E} = \delta_{k, k'} n^{av}_k, \label{eq:simex3}
	\end{eqnarray}
where $n^{av}_k$ is the occupation number of the modes averaged with the coefficients appearing in the decomposition
in Eq.(\ref{eq:opd}), i.e.,
\begin{equation}\label{eq:nav}
n^{av}_k =  \sum_{\alpha=1}^\mathfrak{N} \omega_\alpha \Tr[D_\alpha] n^\alpha_k.
\end{equation}
Replacing Eqs.(\ref{eq:hitex2}) into Eq.(\ref{eq:TCL_Delta_S}) using (\ref{eq:simex2}) and (\ref{eq:simex3})
and taking into account the continuum limit,
we get
\begin{align}
\frac{d}{dt}\rho_S(t)=&\sum_{\alpha=1}^\mathfrak{N} w_\alpha \Bigg\{ -i
[\Delta I_+^\alpha (t)\sigma_+ \sigma_-+ \Delta I_-^\alpha (t) \sigma_- \sigma_+, D_{\alpha}] \notag\\&
+ \Delta R_+^\alpha (t)\mathcal{D}_+[D_\alpha]
+ \Delta R_-^\alpha (t)\mathcal{D}_-[D_\alpha]\Bigg\} \notag\\&
-iI_+^E (t) [\sigma_+\sigma_-, \rho_S(t)] -iI_-^E (t) [\sigma_-\sigma_+, \rho_S(t)]\notag\\&
+ R_+^E (t) \mathcal{D}_+[\rho_S (t)] + R_-^E (t) \mathcal{D}_-[\rho_S (t)],\label{eq:xx1}
\end{align}
where we defined the map
\begin{equation}\label{eq:lali}
	\mathcal{D}_\pm[O]=\sigma_\mp O \sigma_\pm-\frac{1}{2}\{\sigma_\pm \sigma_\mp, O\},
\end{equation}
as well as the functions
\begin{align}
	R_+^\alpha (t)&=\int_0^\infty J(\omega) (n^\alpha(\omega)+1)\frac{\sin[(\varsigma-\omega)t]}{\varsigma-\omega} ,\label{eq:rrii}\\
	R_-^\alpha (t)&=\int_0^\infty J(\omega) n^\alpha(\omega) \frac{\sin[(\varsigma-\omega)t]}{\varsigma-\omega} ,\nonumber\\
	I_+^\alpha (t)&=\int_0^\infty J(\omega) (n^\alpha(\omega)+1)\frac{1-\cos[(\varsigma-\omega)t]}{\varsigma-\omega},\nonumber\\
	I_-^\alpha (t)&=-\int_0^\infty J(\omega)n^\alpha(\omega) \frac{1-\cos[(\varsigma-\omega)t]}{\varsigma-\omega},
	\nonumber\\
	\Delta I_{\pm}^\alpha(t) &= I_\pm^\alpha (t)-I_\pm^E (t), \quad \Delta R_{\pm}^\alpha(t)= R_\pm^\alpha (t)-R_\pm^E (t), \nonumber
\end{align}
and indeed $R_\pm^E(t)$ and $I_\pm^E (t)$ are defined as, respectively, 
$R_\pm^\alpha (t)$ and $I_\pm^\alpha (t)$, but with $n^\alpha(\omega)$ replaced by
$n^{av}(\omega)$.
Interestingly, Eq.(\ref{eq:lali}) shows that both the homogeneous and inhomogeneous parts of the second-order TCL master equation (\ref{eq:xx1})
are written in the canonical form \cite{Gorini1976,Hall2014,Breuer2002}, generalizing the standard
Gorini-Kossakowski-Lindblad-Sudarshan \cite{Gorini1976,Lindblad1976} one to the time-dependent case.

On the other hand, replacing Eqs.(\ref{eq:hitex2}) into 
Eq.(\ref{eq:ref_TCL_Q}) and exploiting again (\ref{eq:simex2}) and (\ref{eq:simex3})  we obtain that the second-order APO description of the dynamics
reads
\begin{align}
\label{eq:adapt_TCL}
\frac{d}{dt}D_{\alpha}(t)=&-i I_+^\alpha (t) [\sigma_+\sigma_-, D_{\alpha}(t)] -i I_-^\alpha (t) [\sigma_-\sigma_+, D_{\alpha}(t)] \notag\\&+  R_+^\alpha (t) \mathcal{D}_+[D_\alpha (t)] + R_-^\alpha (t) \mathcal{D}_-[D_\alpha (t)].
\end{align}
Indeed, we have an uncoupled system of homogeneous equations, each of which takes the canonical
form already mentioned above.
The time-dependent functions defining the master equation are the same real and imaginary parts
of the environmental interaction operators with respect to the states $\rho_\alpha$ fixed by
Eq.(\ref{eq:rrii}) 

In appendix \ref{app:sol}, we report the analytical solutions of Eqs.(\ref{eq:xx1}) and (\ref{eq:adapt_TCL}),
which is at the basis of the comparison between the standard and the APO solutions performed in
the next paragraph.

\subsubsection{Comparison between the two approximated descriptions}
\label{sec:comp-with-inhom}
For the sake of concreteness, we focus also in this case
on initial pure entangled global states and, in particular, we consider states in the form
\begin{equation}\label{eq:initdamp}
	\ket{\Psi}= C_0 \ket{0}\otimes \ket{0} + C_1 \ket{1}\otimes \ket{\{N_k\}_k},
\end{equation}
where $\ket{\{N_k\}_k}$ denotes the pure environmental state with $N_k$ bosons in the mode of frequency $\omega_k$.
The Pauli-decomposition of this state (see Eqs.(\ref{eq:paulib1})-(\ref{eq:paulib3})) is thus fixed by
\begin{equation}
	w_0=w_1=w_2=1~~;~~w_3=2|C_1|^2
\end{equation}
and
\begin{align}
	\rho_0&= |C_0|^2 \ket{0}\bra{0} + |C_1|^2\ket{\{N_k\}_k}\bra{\{N_k\}_k}, \nonumber\\
	\rho_j&= \ket{\xi_j}\bra{\xi_j};
\end{align}
where
\begin{align}
	\ket{\xi_1}&\equiv C_0 \ket{0}+C_1 \ket{\{N_k\}_k} , \nonumber\\
	\ket{\xi_2}&\equiv C_0 \ket{0}+iC_1 \ket{\{N_k\}_k} , \nonumber \\
	\ket{\xi_3}&\equiv\ket{\{N_k\}_k}.
\end{align}
From this, we readily obtain the average numbers of bosons
\begin{eqnarray}
n^0_k&=&n^1_k=n^2_k=|C_1|^2N_k, \nonumber \\
n^3_k&=&N_k,
\end{eqnarray}
and hence the explicit expression of the functions fixing both the standard and the APO second order 
master equations.
Finally, we perform the continuum limit and consider an Ohmic spectral density \cite{Breuer2002}
\begin{equation}\label{eq:flat}
J(\omega) = \gamma \omega \Theta(\omega - \omega_c),
\end{equation}
where $\gamma$ is an adimensional parameter setting the overall strength of the system-environment interaction,
and the Heaviside theta function $\Theta$ introduces a hard cut-off to the maximum value of the frequency $\omega_c$.
Moreover, we consider $N$ bosons for each mode
up to the cut-off frequency $\omega_c$, i.e, (in the continuum limit)
\begin{equation}\label{eq:ndamped}
N(\omega) = N \Theta(\omega - \omega_c).
\end{equation}

\begin{figure}[ht!]
	\centering
	{\bf $\quad$ (a)} \hskip4.1cm{\bf (b)}\\
	\vspace{0.15cm}
		\includegraphics[width=0.48\linewidth]{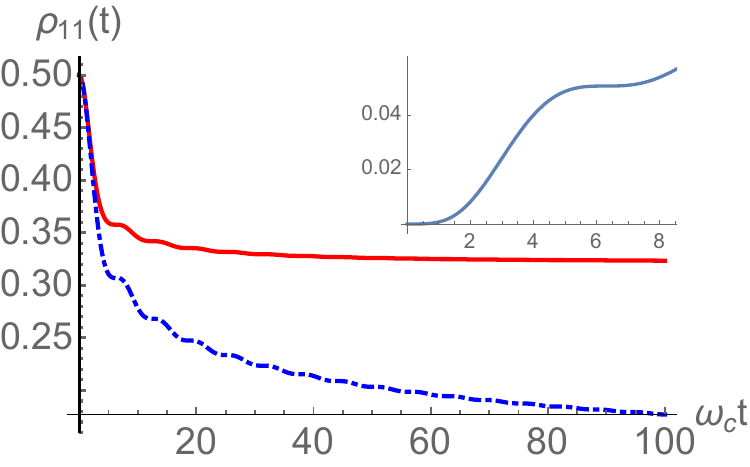}
		\includegraphics[width=0.48\linewidth]{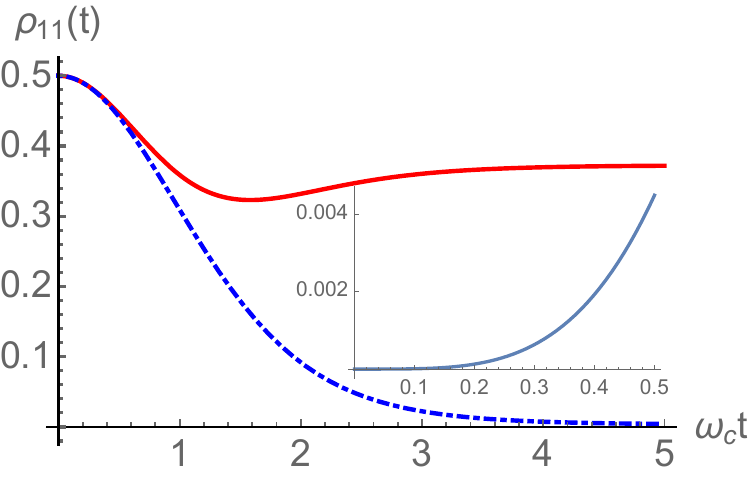}\\
		\vspace{0.35cm}
		{\bf $\quad$ (c)} \hskip4.1cm{\bf (d)} \\	
		\vspace{0.15cm}
		\includegraphics[width=0.48\linewidth]{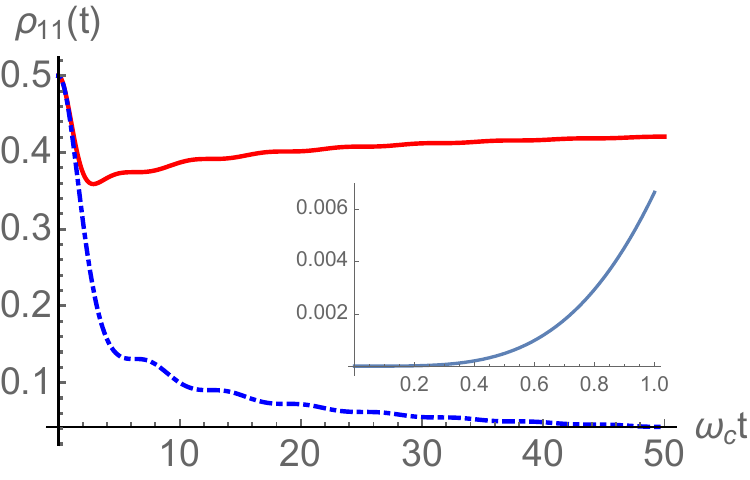}
		\includegraphics[width=0.48\linewidth]{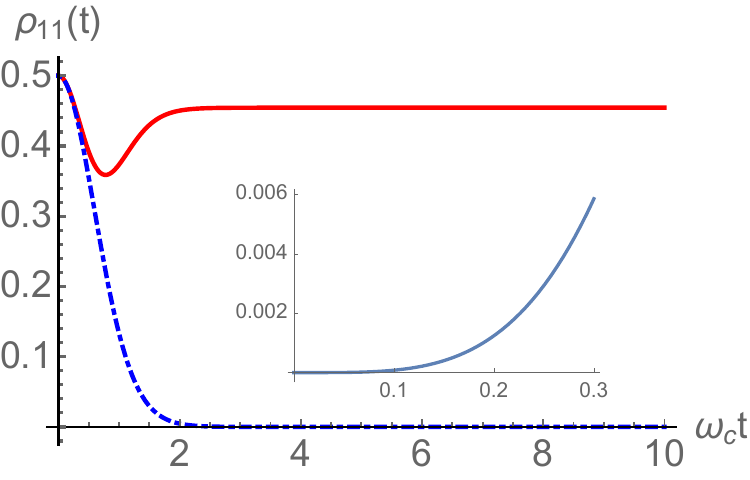}\\

	\caption{Second order APO (red, solid line)
	and second order TCL (blue, dot-dashed line) solutions for $\rho_{11}(t)$ as a function of time $t$,
	for the damped
	two-level system dynamics fixed by Eq.(\ref{eq:hdamp}), with a spectral
	density as in Eq.(\ref{eq:flat}) and an initial
	correlated global state as in Eq.(\ref{eq:initdamp}), with $C_0 = C_1 = 1/\sqrt{2}$
	and number of bosons in the mode with frequency $\omega$ as in Eq.(\ref{eq:ndamped}).
	The 4 panels are referred to different values of $\gamma$ and $N$,
	{\bf (a)} $\gamma = 0.05, N =3$, {\bf (b)} $\gamma=0.5, N = 3$, 
	{\bf (c)} $\gamma = 0.05, N =10$ and {\bf (d)} $\gamma = 0.5, N =10$,
	and the insets report the difference between the APO and TCL second order solution in the short-time regime;
	in all panels, $\omega_c/\nu = 100, \varsigma=0$.}
	\label{fig:6}
\end{figure}

In Fig.\ref{fig:6}, we report the second order solutions of the TCL (blue, dot-dashed line)
and APO (red, solid line) of the excited-state population $\rho_{11}(t)$, for different values of the coupling strength $\gamma$
and number of bosons $N$, for an initial pure 
state as in Eq.(\ref{eq:initdamp}) that is maximally entangled, i.e., for $C_0=C_1=1/\sqrt{2}$;
the coherence $\rho_{10}(t)$ is identically equal to zero at all times.
We observe that the two descriptions agree approximately only in the short-time regime (shown in the insets), while
they depart quite significantly already at intermediate times. Moreover, the difference
between the APO and TCL solutions is enhanced by larger values of the coupling strength and number of bosons.
In any case, also for this model, the two approximations lead to very different predictions
about the asymptotic behavior. In particular, the second order TCL solution always yields
a complete decay to the ground state, 
while the second order APO solution provides us with a finite non-zero asymptotic value
of the excited state population, compatibly with the fact that the two-level system is damped by an environment
that is not in the vacuum state; indeed, the asymptotic value is larger for higher values of the number of bosons $N$ initially in 
the environment, as can be observed by comparing the first and second row of Fig.\ref{fig:6}.

\begin{figure}[ht!]
	\centering
		\includegraphics[width=0.85\linewidth]{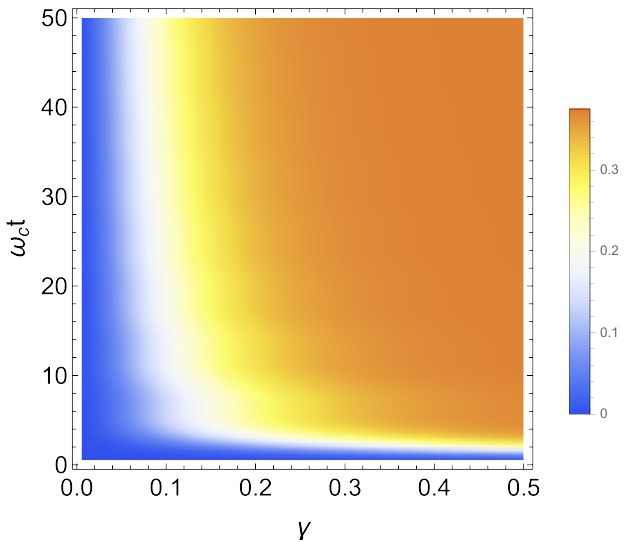}
	\caption{Difference between the second order APO and TCL solutions for $\rho_{11}(t)$, as
	a function of time $t$ and coupling constant $\gamma$, for $N=3$;
	the other parameters are as in Fig.\ref{fig:6}.}
	\label{fig:7}
\end{figure}

The difference between the APO and TCL second order solutions for $\rho_{11}(t)$
is further investigated in Fig.\ref{fig:7}, where it is shown as a function of both time $t$ and coupling 
strength $\gamma$. Again, we see how such a difference is negligible only at short times and/or for weak couplings,
while it leads to different asymptotic values already for intermediate values of the couplings.
In addition, we note some oscillations in time of the difference between the APO and TCL solution 
(also observable in Fig.\ref{fig:6} {\bf (a)}), which are suppressed by larger values of the coupling.

Let us stress that in our awareness this is the first treatment of initial correlations between system and bath in this model.

\section{Conclusions and outlook}
\label{sec:conclusions-outlook}
We have developed a perturbative approach for the treatment of open quantum system dynamics 
that is able to deal with general microscopic models of the system-environment interaction and, above all, with arbitrary,
possibly correlated initial global states. Our approach combines features of the standard projection operator techniques
with a convenient decomposition of the initial state obtained relying on frame-theory. The initial state is expressed as a convex combination of product operators,
which involve proper states on the environmental side and whose number is limited by the square dimension of the open system.
As a result, the dynamics of the open system is characterized by a limited set of differential equations uncoupled and homogeneous even for correlated initial states,
at variance with existing techniques. This has allowed us to deal with correlated initial states in a spin-boson scenario.
The equations are fixed by environmental correlation functions with a clear physical meaning, which generalize the usual covariance
functions and can be in principle accessed experimentally. 
The detailed analysis of two significant two-level system dynamics, i.e., pure dephasing and damping by a continuous bosonic bath,
also shows that 
our method reproduces expected dynamical behaviors in the long-time regime
more closely than the standard approach.

To further appreciate  the
potential and versatility of our method,
it will be important to take into account
more complex open-system dynamics, and a first step in this direction might be the study of multi-qubit evolutions where the mentioned decomposition
of the initial global state has been already applied successfully \cite{Raja2020}. In addition, 
the effectiveness of the projection-operator approach we introduced here will be clarified by a systematic analysis of higher-order
contributions, as well as by the analogous treatment for the time-non-local form of the equations of motion, which can give an improved approximation of the dynamics in certain circumstances \cite{Breuer2004,Reimer2019b}.
Finally, a realistic treatment of the correlations between an open quantum system and its environment  
at the initial time will help reach a
full understanding of the connection between the (quantum or classical) system-environment correlations and their impact on the subsequent dynamics.


\acknowledgments
All authors acknowledge
support from UniMi, via Transition Grant
H2020 and PSR-2 2020. NM acknowledges funding
by the Alexander von Humboldt Foundation in the
form of a Feodor-Lynen Fellowship.

\onecolumngrid
\appendix
\section{Second-order master equation for a correlated state projection}\label{app:a}
In this section, we give a more explicit, albeit unavoidably cumbersome, expression for the second order master
equation (\ref{eq:dyn_Delta_corr_proj2}), obtained by combining a generic correlated-state projector and the decomposition
of the initial global state as in Eq.(\ref{eq:opd}).

Using the definitions in Eq.(\ref{eq:jjcorr}),
Eq.(\ref{eq:dyn_Delta_corr_proj2}) can be written as
\begin{eqnarray}
\label{eq:TCL_Delta_Scorr}
\frac{d}{dt}\eta_i(t)&=&\sum_{\alpha=1}^{\mathfrak{N}}\omega_{\alpha} \Bigg[ -ig
\sum_j \Big(A_j(t) D_\alpha  \braket{\overline{Y}_i B_j(t)}_{\tilde{\Delta}_\alpha} - D_\alpha A_j(t)\braket{ B_j(t)\overline{Y}_i}_{\tilde{\Delta}_\alpha}\Big)
 \notag\\&&
 - g^2\sum_{j_1,j_2}\int_0^t d \tau
	\Bigg(
	A_{j_1}(t)A_{j_2}(\tau)D_\alpha  \mathfrak{H}^{(\tilde{\Delta}_\alpha)}_{i; j_1 j_2}(t,\tau) 
	-A_{j_1}(t)D_\alpha A_{j_2}(\tau) \mathfrak{K}^{(\tilde{\Delta}_\alpha)}_{i; j_2 j_1}(\tau,t) \\&&
	-A_{j_2}(\tau)D_\alpha A_{j_1}(t)\mathfrak{L}^{(\tilde{\Delta}_\alpha)}_{i; j_1 j_2}(t,\tau)
	+D_\alpha  A_{j_2}(\tau)A_{j_1}(t)\mathfrak{M}^{(\tilde{\Delta}_\alpha)}_{i; j_2 j_1}(\tau,t)\Bigg)
	\Bigg]\notag\\
&&-ig\sum_{j,j_1}\Bigg(A_{j_1}(t) \eta_j(t)  \braket{\overline{Y}_i B_{j_1}(t)}_{\overline{X}_j} - \eta_j(t) A_{j_1}(t)\braket{B_{j_1}(t)\overline{Y}_i}_{\overline{X}_j}\Bigg) \notag\\&& 
 - g^2\sum_{j,j_1,j_2}\int_0^t d \tau 
	\Bigg(
	A_{j_1}(t)A_{j_2}(\tau)\eta_j(t)  \mathfrak{H}^{(\overline{X}_j)}_{i; j_1 j_2}(t,\tau)
	-A_{j_1}(t)\eta_j(t) A_{j_2}(\tau)\mathfrak{K}^{(\overline{X}_j)}_{i; j_2 j_1}(\tau,t)\notag\\&&
	-A_{j_2}(\tau)\eta_j(t) A_{j_1}(t)\mathfrak{L}^{(\overline{X}_j)}_{i; j_1 j_2}(t,\tau)
	+\eta_j(t)  A_{j_2}(\tau)A_{j_1}(t)\mathfrak{M}^{(\overline{X}_j)}_{i; j_2 j_1}(\tau,t)\Bigg)
	\Bigg], \notag
\end{eqnarray}
where we introduced the functions (implying their dependence
on the environmental operators $\{\overline{Y}_i\}$ and $\{\overline{X}_i\}$)
\begin{eqnarray}
\mathfrak{H}^{(O)}_{i; j_1 j_2}(t,\tau) &=& 
\braket{\overline{Y}_{i} B_{j_1}(t)B_{j_2}(\tau)}_{O}  \label{eq:funfun}\\
&&-\sum_{i_0}\braket{\overline{Y}_{i_0}B_{j_2}(\tau)}_{O} \braket{\overline{Y}_{i}B_{j_1}(t)}_{\overline{X}_{i_0}} \notag\\
\mathfrak{K}^{(O)}_{i; j_2 j_1}(\tau,t)&=& \braket{B_{j_2}(\tau)\overline{Y}_{i} B_{j_1}(t)}_{O} \notag\\
&&- \sum_{i_0}\braket{B_{j_2}(\tau)\overline{Y}_{i_0}}_{O} \braket{\overline{Y}_{i}B_{j_1}(t)}_{\overline{X}_{i_0}} \notag\\
\mathfrak{L}^{(O)}_{i; j_1 j_2}(t,\tau) &=& \braket{B_{j_1}(t)\overline{Y}_{i} B_{j_2}(\tau)}_{\tilde{\Delta}_\alpha}\notag\\
&&-\sum_{i_0} \braket{\overline{Y}_{i_0}B_{j_2}(\tau)}_{\tilde{\Delta}_\alpha} \braket{B_{j_1}(t)\overline{Y}_{i}}_{\overline{X}_{i_0}} \notag\\
\mathfrak{M}^{(O)}_{i; j_2 j_1}(\tau,t)&=&\braket{B_{j_2}(\tau) B_{j_1}(t)\overline{Y}_{i}}_{\tilde{\Delta}_\alpha}\notag\\
&&-\sum_{i_0} \braket{B_{j_2}(\tau)\overline{Y}_{i_0}}_{\tilde{\Delta}_\alpha} \braket{B_{j_1}(t)\overline{Y}_{i}}_{\overline{X}_{i_0}}.\notag
\end{eqnarray}
We note that the presence of the operators 
$\{\overline{X}_i\}$ and $\{\overline{Y}_i\}$ related with it does not allow us to express
the terms in the equation by means of (generalized) correlation
functions of the environmental interaction operators as done with Eqs.(\ref{eq:notagf}), (\ref{eq:notagg}) and (\ref{eq:cov}), but the more general functions in Eq.(\ref{eq:funfun}) are needed.

\section{Analytic solutions of the second-order master equations for the damped two-level system}\label{app:sol}
Here we provide the explicit analytic solutions of Eqs.(\ref{eq:xx1}) and (\ref{eq:adapt_TCL}), which correspond
to the second-order description of the dynamics of a two-level open system damped by a bosonic bath according to,
respectively, the standard and the APO perturbative expansions.

Introducing the functions
\begin{align}
	\bar{R}_\alpha(t)&=R_+^\alpha (t)+R_-^\alpha (t)=\int_0^\infty J(\omega) (2n^\alpha(\omega)+1) \frac{\sin[(\varsigma-\omega)t]}{\varsigma-\omega} , \nonumber\\
	\bar{I}_\alpha(t)&=I_+^\alpha (t)-I_-^\alpha (t)=\int_0^\infty J(\omega) 
(2n^\alpha(\omega)^\alpha+1) \frac{1-\cos[(\varsigma-\omega)t]}{\varsigma-\omega},\label{eq:extw1}
\end{align}
as well as 
$\bar{R}_E(t)=R^E_+(t)+R^E_- (t)$ and $\bar{I}_E(t)=I^E_-(t)+I^E_- (t)$
and using
\begin{align}
	\Tr\{D_{\alpha}(t)\}&=\Tr\{D_{\alpha}\}, \quad
	\braket{0|D_{\alpha}(t)|1}=\braket{1|D_{\alpha}(t)|0}^*,\notag\\
	\braket{1|\mathcal{D}_+[D]|1}&=-\braket{1|D|1},\quad
	\braket{1|\mathcal{D}_-[D]|1}=\Tr\{D\}-\braket{1|D|1},\notag\\
	\braket{1|\mathcal{D}_+[D]|0}&=-\frac{1}{2}\braket{1|D|0}, \quad
	\braket{1|\mathcal{D}_-[D]|0}=-\frac{1}{2}\braket{1|D|0}, \label{eq:extw2}
\end{align}
Eq.(\ref{eq:xx1}) leads to
\begin{align}
\frac{d}{dt}\rho_{11}(t)=&{\mu}(t)
-\bar{R}_E (t) \rho_{11}(t) ,\nonumber\\
\frac{d}{dt}\rho_{10}(t)=&\nu(t)
-\frac{1}{2}\Big(i\bar{I}_E (t)+\bar{R}_E (t)\Big) \varrho_{10}(t),\label{eq:xxapp}
\end{align}
where
\begin{align}
	\mu(t)&
	=\sum_{\alpha=1}^\mathfrak{N} w_\alpha \Bigg\{-\bar{R}_\alpha (t)\braket{1|D_{\alpha}|1} + R_-^\alpha (t)\Tr\{D_\alpha\}\Bigg\}+ \varrho_{11}(0)\bar{R}_E (t), \nonumber
	\\
	\nu(t)&=-\frac{1}{2}\sum_{\alpha=1}^\mathfrak{N} w_\alpha\braket{1|D_\alpha|0} \Bigg\{ i\Big(\bar{I}_\alpha (t)-\bar{I}_E (t)\Big) 
	+ \Big(\bar{R}_\alpha (t)-\bar{R}_E (t)\Big)
	\Bigg\}.
\end{align}
The solutions to Eq.(\ref{eq:xxapp}) are 
\begin{eqnarray}
	\rho^{TCL}_{11}(t)&=&\rho_{11}(0)\exp\left\{-\int_0^t ds \bar{R}_E (s) \right\}+\int_0^t d\tau \exp\left\{-\int_\tau^t ds \bar{R}_E (s) \right\}{\mu}(\tau), \nonumber\\
	\rho^{TCL}_{10}(t)&=&
	\rho_{10}(0)\exp\left\{-\frac{1}{2}\int_\tau^t ds \Big(i\bar{I}_E (s)+\bar{R}_E (s)\Big)\right\}+\int_0^t d\tau \exp\left\{-\frac{1}{2}\int_\tau^t ds \Big(i\bar{I}_E (s)+\bar{R}_E (s)\Big) \right\}\nu(\tau) .
\end{eqnarray}

On the other hand, using Eqs.(\ref{eq:extw1}) and (\ref{eq:extw2}),
Eq.(\ref{eq:adapt_TCL}) leads to two independent closed differential equations
\begin{eqnarray}
	\frac{d}{dt}\braket{1|D_{\alpha}(t)|1}&=& \Tr\{D_{\alpha}\}R_-^\alpha(t) -\braket{1|D_{\alpha}(t)|1}\bar{R}_\alpha(t), \nonumber\\
	\frac{d}{dt}\braket{1|D_{\alpha}(t)|0}&=&-\frac{1}{2} \Big(-i\bar{I}_\alpha(t)+\bar{R}_\alpha(t)\Big)\braket{1|D_{\alpha}(t)|0},
\end{eqnarray}
whose solutions read
\begin{eqnarray}
	\braket{1|D_{\alpha}(t)|1}&=&
	\braket{1|D_{\alpha}|1}\exp\left\{-\int_0^t ds \bar{R}_\alpha(s) \right\}+\Tr\{D_{\alpha}\}\int_0^t d\tau \exp\left\{-\int_{\tau}^t ds \bar{R}_\alpha(s) \right\}R_-^\alpha(\tau) \nonumber\\
	\braket{1|D_{\alpha}(t)|0}&=&\braket{1|D_{\alpha}|0} \exp\left\{-\frac{1}{2}\int_0^t d\tau i\bar{I}_\alpha(\tau)+\bar{R}_\alpha(\tau)\right\}.
\end{eqnarray}
Notice that choosing the Pauli decomposition we have that $D_0$ is the only operator with non-vanishing trace and $\braket{1|D_0|1}=0$, so that we get
\begin{align}
\rho_{11}^{APO}(t)&=\sum_\alpha w_\alpha\braket{1|D_{\alpha}(t)|1}\notag\\&
=\rho_{11}(0)\exp\left\{-\int_0^t ds \bar{R}_3(s) \right\}+\int_0^t d\tau \exp\left\{-\int_{\tau}^t ds \bar{R}_0(s) \right\}R_-^0(\tau),\\
\rho_{10}^{APO}(t)=&\sum_\alpha w_\alpha\braket{1|D_{\alpha}(t)|0}\notag\\= &
-\frac{1}{2}(1-i)\exp\left\{-i\frac{1}{2}\int_0^t d\tau\bar{I}_0(\tau)\right\}\exp\left\{-\frac{1}{2}\int_0^t d\tau \bar{R}_0(\tau)\right\}\notag\\&
+\frac{1}{2}w_1\exp\left\{-i\frac{1}{2}\int_0^t d\tau\bar{I}_1(\tau)\right\}\exp\left\{-\frac{1}{2}\int_0^t d\tau \bar{R}_1(\tau)\right\}\notag\\&
-\frac{1}{2}i w_2\exp\left\{-i\frac{1}{2}\int_0^t d\tau\bar{I}_2(\tau)\right\}\exp\left\{-\frac{1}{2}\int_0^t d\tau \bar{R}_2(\tau)\right\}. &
\end{align}

\end{document}